\documentclass[12pt]{article}

\usepackage{graphicx,url}
\usepackage[english]{babel} 
\usepackage[utf8]{inputenc}  
\usepackage{verbatim}
\usepackage{listings}
\usepackage{indentfirst}
\usepackage{xcolor}
\usepackage{amsmath}
\usepackage{bm}
\usepackage{amsfonts}
\usepackage{amssymb}
\usepackage{caption}
\usepackage{ulem}
\usepackage{float}
\usepackage{multirow}

\usepackage{titlesec}
\titleformat{\paragraph}[block]{\normalfont\normalsize\bfseries}{\theparagraph}{1em}{}
\titlespacing*{\paragraph}{0pt}{1em}{1em}

\title{\textbf{FDTRImageEnhancer: Combining Physics-Informed Deconvolution and Microstructure-Aware Deep Learning to Enhance Thermal Images}}

\author{
    Alesanmi R. R. Odufisan\textsuperscript{1}\thanks{Corresponding author: \texttt{alesanmiodufisan@gmail.com}}
}

\date{} 

\begin{document} 
\maketitle

\begin{center}
    \footnotesize 
    \textsuperscript{1}Department of Theoretical and Applied Mechanics, Northwestern University \\
\end{center}

\begin{abstract}
\footnotesize

We present \textit{FDTRImageEnhancer}, an open-source computational framework that integrates physics-informed deconvolution with microstructure-aware deep learning for inverse thermal property mapping. The approach is demonstrated on Frequency Domain ThermoReflectance (FDTR) phase data, but its architecture is general to problems where high-resolution structural information must be reconciled with lower-resolution physical measurements.

Drawing inspiration from continuum damage mechanics, where sharp cracks are represented in a smeared or nonlocal form, we approximate FDTR’s spatial averaging as a two-parameter Gaussian convolution mimicking pump and probe laser profiles. This physics abstraction is coupled with a neural network that infers region-specific thermal conductivity, guided by structural segmentation via k-means clustering to reduce the parameter space.

As a methodological proof of concept, the framework was tested using synthetic FDTR data generated from finite element simulations. Across multiple runs, it consistently identified reduced thermal conductivity at grain boundaries (GBs) that were visually indistinguishable in analytically inverted, lower-resolution conductivity maps. While bulk conductivity values were recovered with high accuracy ($<0.5\%$ error), the model overestimated the GB conductivity reduction - likely due to the simplified treatment of FDTR physics in the Gaussian abstraction. Through computational optimization, the total runtime was reduced to just a few minutes, demonstrating its potential utility in practical cases. Overall, this proof of concept establishes a foundation for future convolutional neural network (CNN)–based models trained in a supervised manner on diverse FDTR datasets, capable of reproducing full FDTR physics almost instantaneously.

The full Python implementation, including example datasets, is provided to enable complete reproducibility and adaptation to other inverse thermal modeling tasks.


\end{abstract}

\newpage
\section{Introduction}

The ability to predict and measure how heat flows through materials is central to many areas of science and engineering, ranging from improving microelectronic cooling to designing more efficient energy systems. In recent years, advances in computational physics and artificial intelligence (AI) have begun to transform thermal modeling, revealing subtle relationships between a material's structure and its thermal properties that are often difficult to capture with conventional methods.

For example, machine-learned interatomic potentials (MLIPs) have been incorporated into molecular dynamics simulations to improve force predictions beyond the accuracy of empirical potentials \cite{fujii2022, li2020}. These data-driven models allow for more faithful representations of atomic interactions, particularly in complex microstructures such as grain boundaries (GBs), where traditional approaches often fall short. Other studies have used machine learning to correlate GB energy with local temperature distributions in disordered systems \cite{fomin2024}, or to directly map experimental measurements to thermal properties with improved speed and robustness compared to traditional fitting \cite{chatterjee2024, shen2020}.

Despite this progress, a persistent challenge remains: the resolution gap between thermal imaging and structural characterization. Thermal measurement techniques such as Frequency Domain ThermoReflectance (FDTR) can achieve micron-scale resolution \cite{isotta2023, isotta2024}, whereas structural imaging methods like electron backscatter diffraction (EBSD) can resolve features down to $\sim$10~nm \cite{geiss2013}. This disparity makes it difficult to directly connect microstructural features - such as grain boundaries - to their influence on thermal transport, especially when those features are visually obscured in the thermal images.  

In this work, we address that gap through \textit{FDTRImageEnhancer}, an open-source computational framework that combines a physics-based Gaussian convolution abstraction of FDTR with microstructure-aware deep learning. While FDTR is our demonstration case, the framework is broadly applicable to problems where high-resolution structural information must be reconciled with lower-resolution physical measurements.  

Our approach draws on the principles of physics-informed neural networks (PINNs), which embed physical laws directly into a network's training process to ensure that solutions remain consistent with both observed data and governing equations \cite{raissi2019, cai2021, zhou2023, qian2023, sripada2024}. Here, we adapt this concept to a methodological framework where the traditional PDE-based physics loss is replaced by a surrogate loss based on discrepancies between network-predicted parameters and physics-based analytical inversion results - a necessary modification due to discontinuities in FDTR's governing equations. The Gaussian convolution captures FDTR's inherent spatial averaging, enabling a microstructure-aware inversion process.

To support adoption and adaptation, this paper follows a tutorial structure: beginning with a one-dimensional PINN example for the heat equation to introduce the core concepts, then presenting the mathematical framework and full FDTR-specific implementation. Since this methodology precomputes the FDTR physics before using it as a surrogate loss, the standard PINN framework is reviewed first to differentiate the two approaches. This design allows researchers - including those new to machine learning - to reproduce our results and extend the method to other inverse problems in thermal transport.

\section{Physics-Informed Neural Networks for the 1D Heat Equation: Recovering Thermal Diffusivity}
\label{sec:1D_tutor}

\begin{figure}[H]
\centering
\includegraphics[scale=0.6]{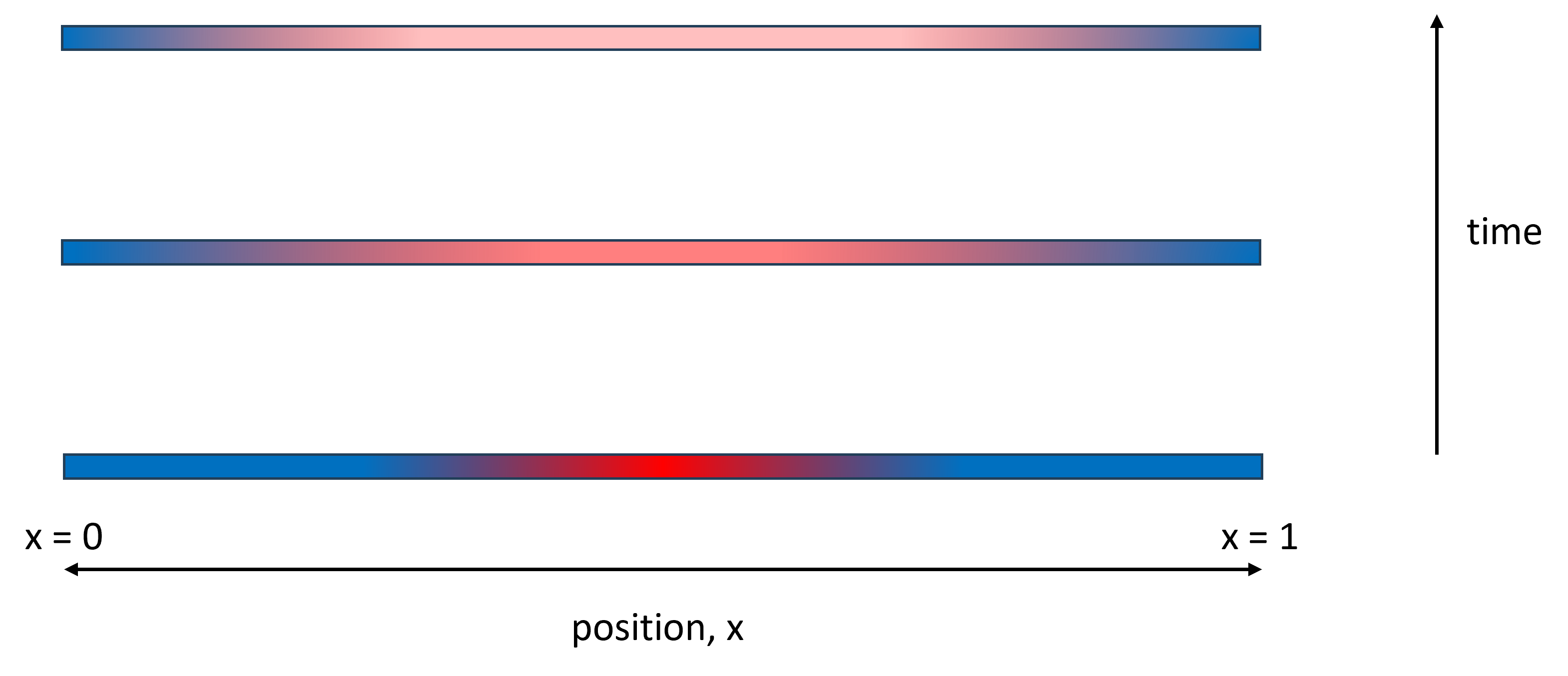}
\caption{1D Bar initially heated at the middle}
\label{fig:1d_bar}
\end{figure}

Consider a metal rod with both ends held at a fixed temperature of 0, and an initial heat concentration that peaks in the center, as shown in Figure \ref{fig:1d_bar}.

This setup can be described by the 1D heat equation, i.e,

\begin{equation}
    \frac{\partial T}{\partial t} = \alpha \frac{\partial^2T}{\partial^2 x}
    \label{eq:1d_fourier}
\end{equation}

Where :

\begin{itemize}
    \item $T(x,t)$: temperature as a function of space and time
    \item $\alpha$: thermal diffusivity
    \item $x \in [0,1],~~~~ t > 0$
\end{itemize}

With the system constrained by the following boundary conditions:

\begin{itemize}
    \item $T(0,t) = T(1,t) = 0$
    \item $T(x,0) = \sin{(\pi x)}$
\end{itemize}

To simulate temperature readings at $50$ points in between $x = 0$ and $x = 1$, the analytical solution was used, i.e:

\begin{equation}
    T(x,t) = \sin{(\pi x)} \cdot \exp{(-\alpha \pi^2 t)}
    \label{eq:1d_fourier_analytical}
\end{equation}

\begin{figure}[H]
    \centering
    \begin{tabular}{ccc}
    
        \includegraphics[width=0.5\textwidth]{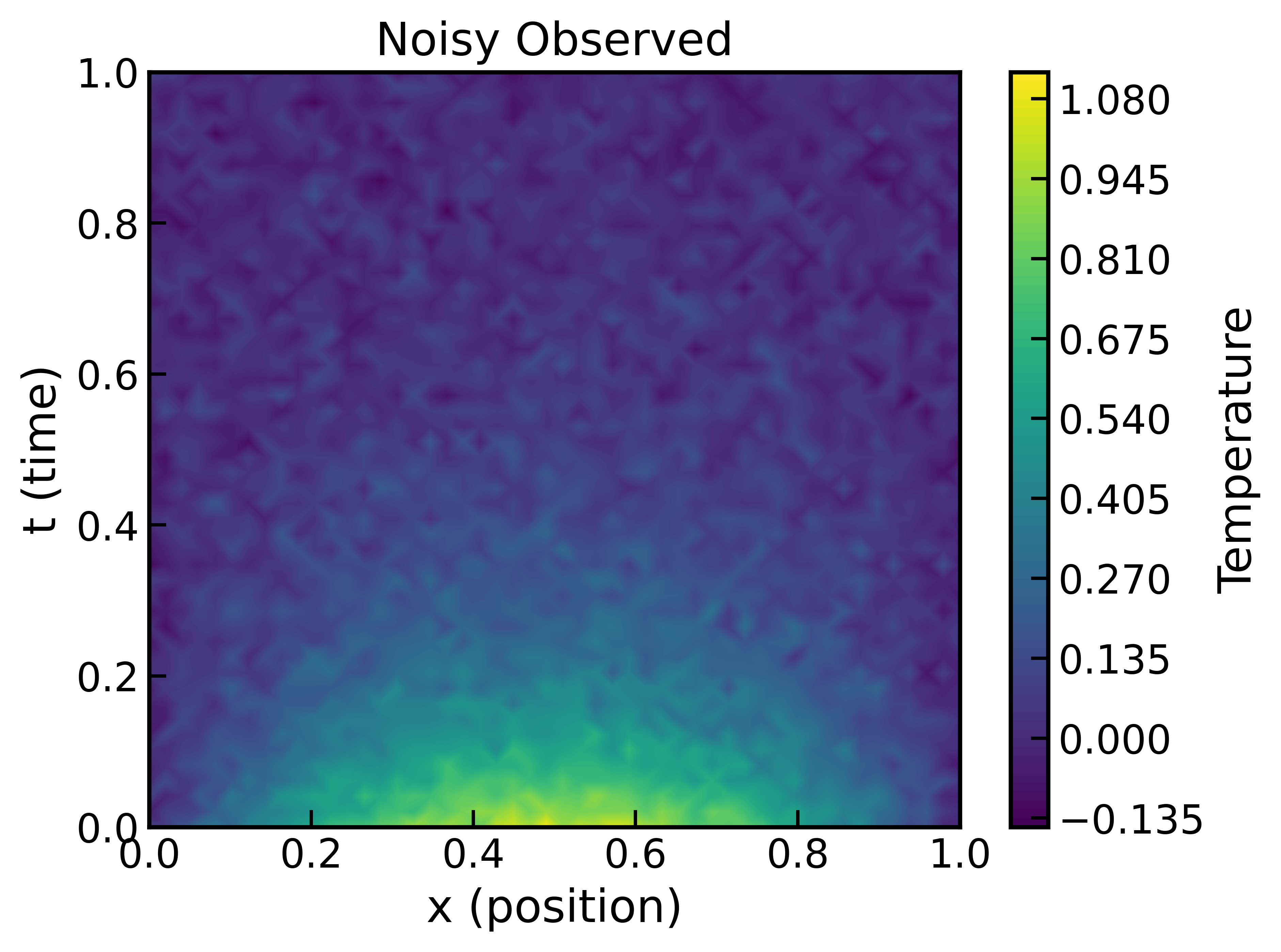} & 
        \includegraphics[width=0.45\textwidth]{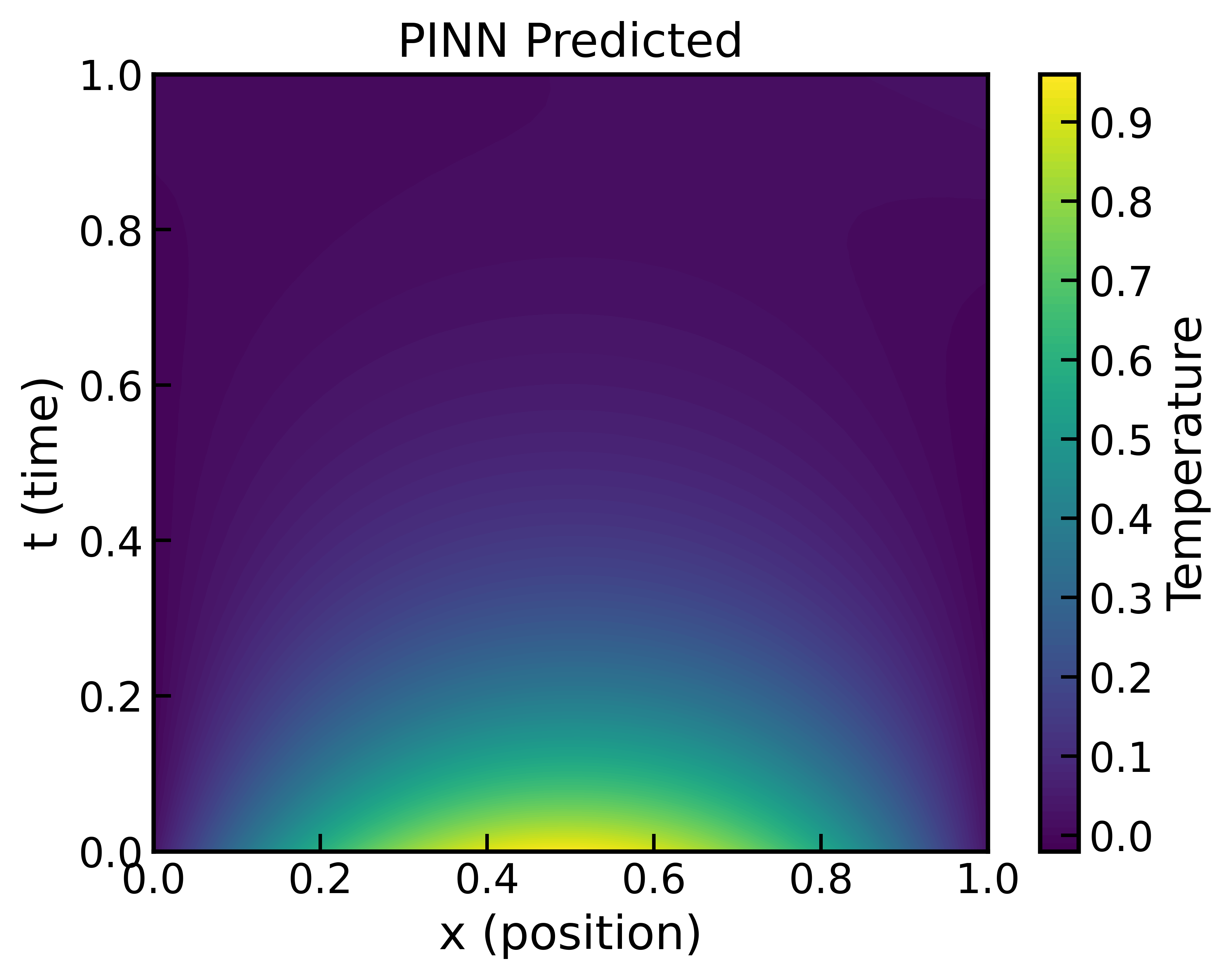} \\ 

        (a) & 
        (b) \\ 
        
    \end{tabular}
    \caption{Temperature profile evolution comparison}
    \label{fig:pinncompare1d}
\end{figure}

For realism, and to demonstrate the effectiveness of the PINN, artificial noise was added to the measurements as shown in Figure \ref{fig:pinncompare1d}(a).

We then define a fully connected feedforward neural network which serves as a ``black-box" function to map spatial and temporal values to temperature with the diffusivity as a learnable parameter, i.e.:

\begin{equation}
    f(x,t; \alpha) \to T
    \label{eq:1d_nn}
\end{equation}

\begin{figure}[H]
\centering
\includegraphics[scale=0.56]{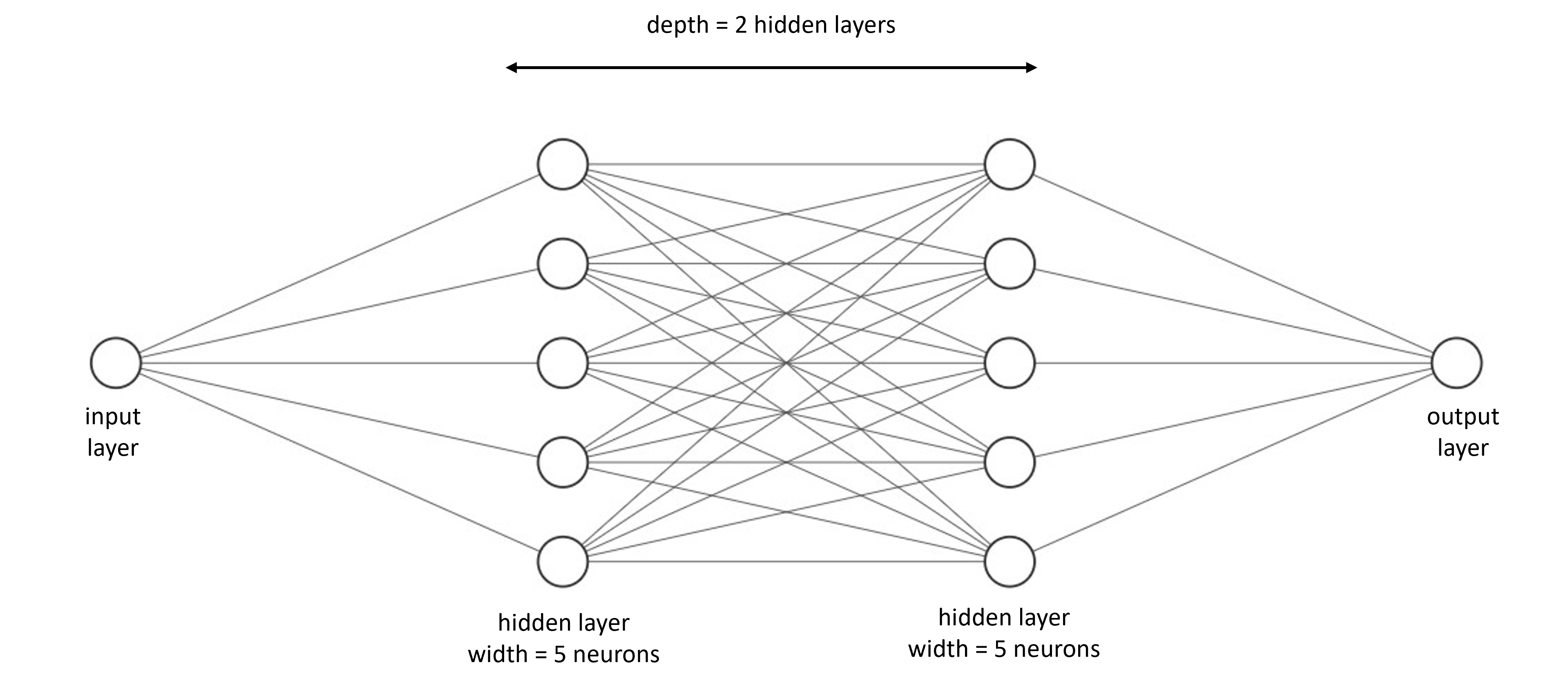}
\caption{Neural network schematic}
\label{fig:nn_schem}
\end{figure}

A ``feedforward" network in this context refers to a neural network where every neuron in one layer is connected to every neuron in the next layer as in Figure \ref{fig:nn_schem}, and information flows in one direction only, i.e. input $\to$ output. In general, increasing the depth of a neural network tends to enhance its ability to capture fine-grained, local features, while increasing the width of each layer improves its capacity to model broader, global patterns \cite{mori2020}. The internal weights and biases of the model are usually handled automatically by commercial and open-source frameworks. For this model, a neural network with two fully connected hidden layers with $64$ neurons each was found to be sufficient.

In neural networks, activation functions introduce nonlinearity - without them it would be equivalent to a multilayered linear interpolation operation. In this model we used the hyperbolic tangent ($\tanh$) activation due to its zero-centered output, smooth differentiability, and proven stability in gradient-based optimization, particularly in problems with low to moderate depth and physics-informed loss functions \cite{lecun2012}.

With the physics of the 1D heat equation embedded in the neural network, the total loss, i.e., the objective to minimize, becomes:

\begin{equation}
    \mathcal{L}_{PINN}(\alpha) = \underbrace{\frac{1}{N} \sum^{N}_{i = 1} (T^{pred}_i(x,t) - T^{obs}_i)^2}_{\text{Data Loss from Neural Network Prediction}} + \underbrace{\frac{1}{N} \sum^{N}_{i = 1} \left(\frac{\partial T^{pred}_i(x,t)}{\partial t} - \alpha \frac{\partial^2T^{pred}_i(x,t)}{\partial^2 x}\right)^2}_{\text{Physics Loss}}
    \label{eq:1d_pinn_loss}
\end{equation}

Where $N$ refers to the number of data points. That is, the neural network attempts to learn how $\alpha$, the diffusivity, affects the mapping from $(x,t)$ to $T$, calculating the Mean Squared Error (MSE) between its prediction and the observed data as its ``data loss". Since this mapping is based solely on patterns in the data, the predicted $T$ and learned $\alpha$ are then substituted into the heat equation to compute the ``physics loss", ensuring that the neural network's predictions are grounded in reality. Figure \ref{fig:pinncompare1d} shows a comparison between the raw data and the PINN-predicted temperature profile evolution.

Although regular least squares fitting could have been used to achieve the same result, it is relevant to note that the PINN needed no knowledge of initial and boundary conditions - the physics loss ensured that predictions made by the PINN were physically realistic.

In the next section, we introduce the FDTR neural network where the physics loss is replaced by a surrogate parameter loss, which would be akin to replacing the physics loss in the 1D example above by a comparison between a network-predicted and analytically computed $\alpha$.

\section{Physics-Informed Gaussian Convolution and Microstructure-Aware Neural Network for FDTR}

\subsection{FDTR Setup}
\begin{figure}[H]
\centering
\includegraphics[scale=0.7]{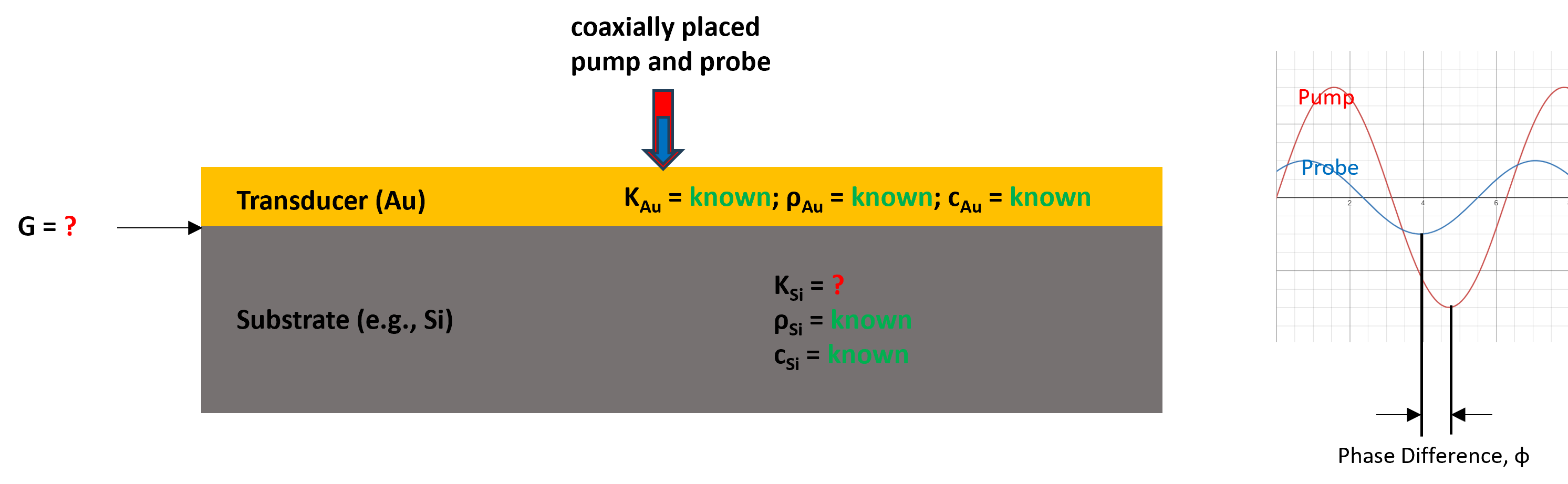}
\caption{FDTR Experiment Setup}
\label{fig:fdtr_setup}
\end{figure}

The Frequency Domain ThermoReflectance (FDTR) technique is a non-destructive optical technique used to obtain the thermal properties of a sample. It works by correlating the phase shift between a sinusoidally modulated pump laser, or heating source, and the corresponding induced temperature increase read by a probe/sensing laser. Figure \ref{fig:fdtr_setup} illustrates this setup.

In this technique, the thermal conductivity of the substrate, $\kappa$, and the interface conductance, $G$ are usually the only unknowns, and they are obtained by minimizing the measured phase and the phase from an analytical model \cite{schmidt2008}, where the phase is given as: 

\begin{equation}
    \phi^{analytical} = \tan^{-1} \left( \frac{\Im[\hat{H}(\omega)]}{\Re[\hat{H}(\omega)]} \right)
    \label{eq:phase}
\end{equation}

The term $\hat{H}(\omega)$ is the complex-valued temperature response, which is expressed as:

\begin{equation}
    \hat{H}(\omega) = \int_0^\infty  \bar{\hat{\mathsf{G}}}(k, \omega)  \bar{\hat{P}}(k, \omega) \bar{\hat{S}}(k, \omega)k ~dk
    \label{eq:h_thermal_response}
\end{equation}

where $\bar{\hat{\mathsf{G}}}(k, \omega)$ is the Hankel transform of the Green’s function solution
for the temperature response due to an oscillating the heating source, $\bar{\hat{P}}(k, \omega)$ (with $\omega = 2\pi f$, $f$ being the frequency of the oscillation), and $\bar{\hat{S}}(k, \omega)$ refers to the probe, or sensing laser. In this expression, $~\hat{}~$ and $~\bar{}~$ indicators refer to expressions in Fourier and Hankel space respectively.

\begin{figure}[H]
    \centering
    \begin{tabular}{ccc}
    
        \includegraphics[width=0.45\textwidth]{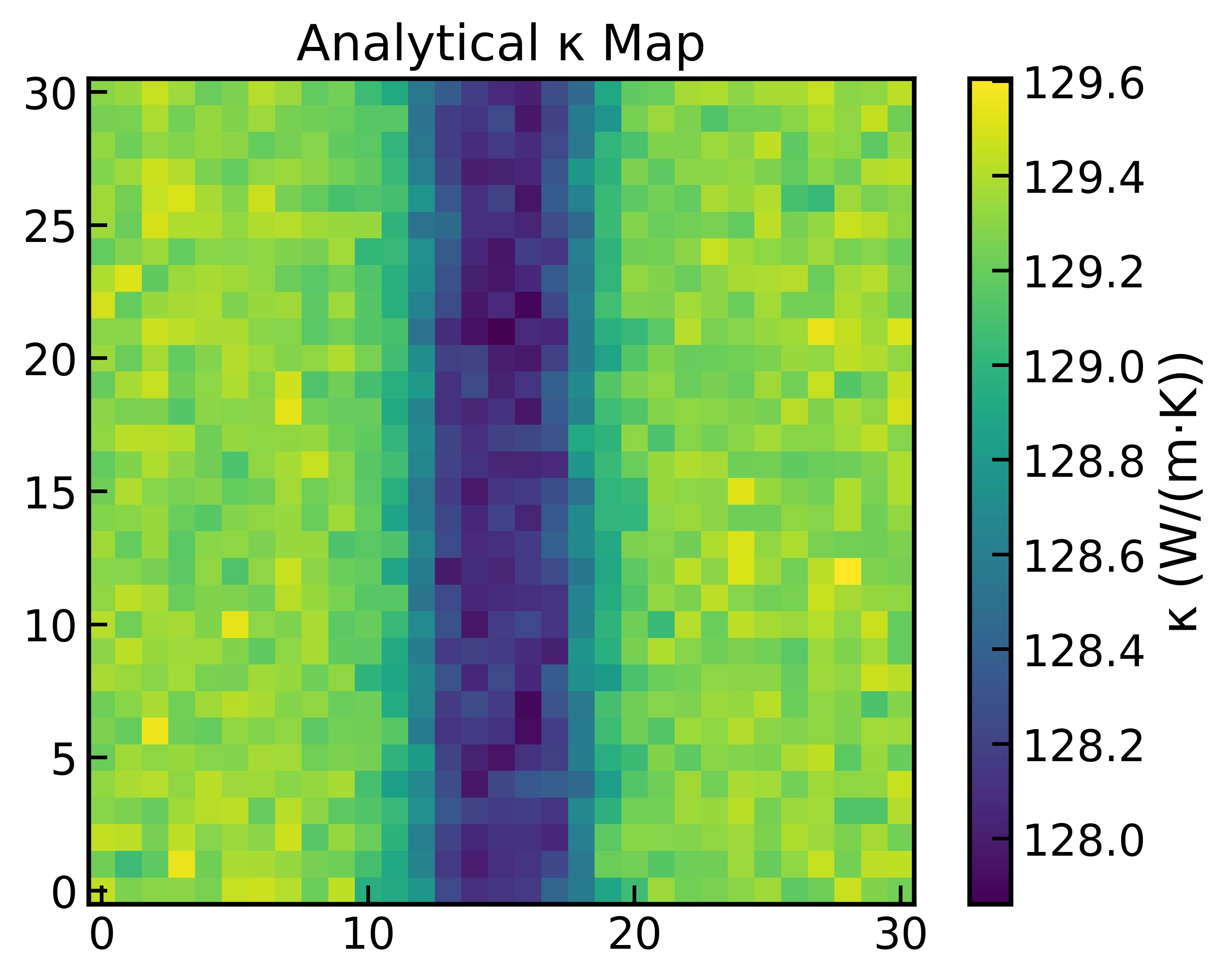} & 
        \includegraphics[width=0.45\textwidth]{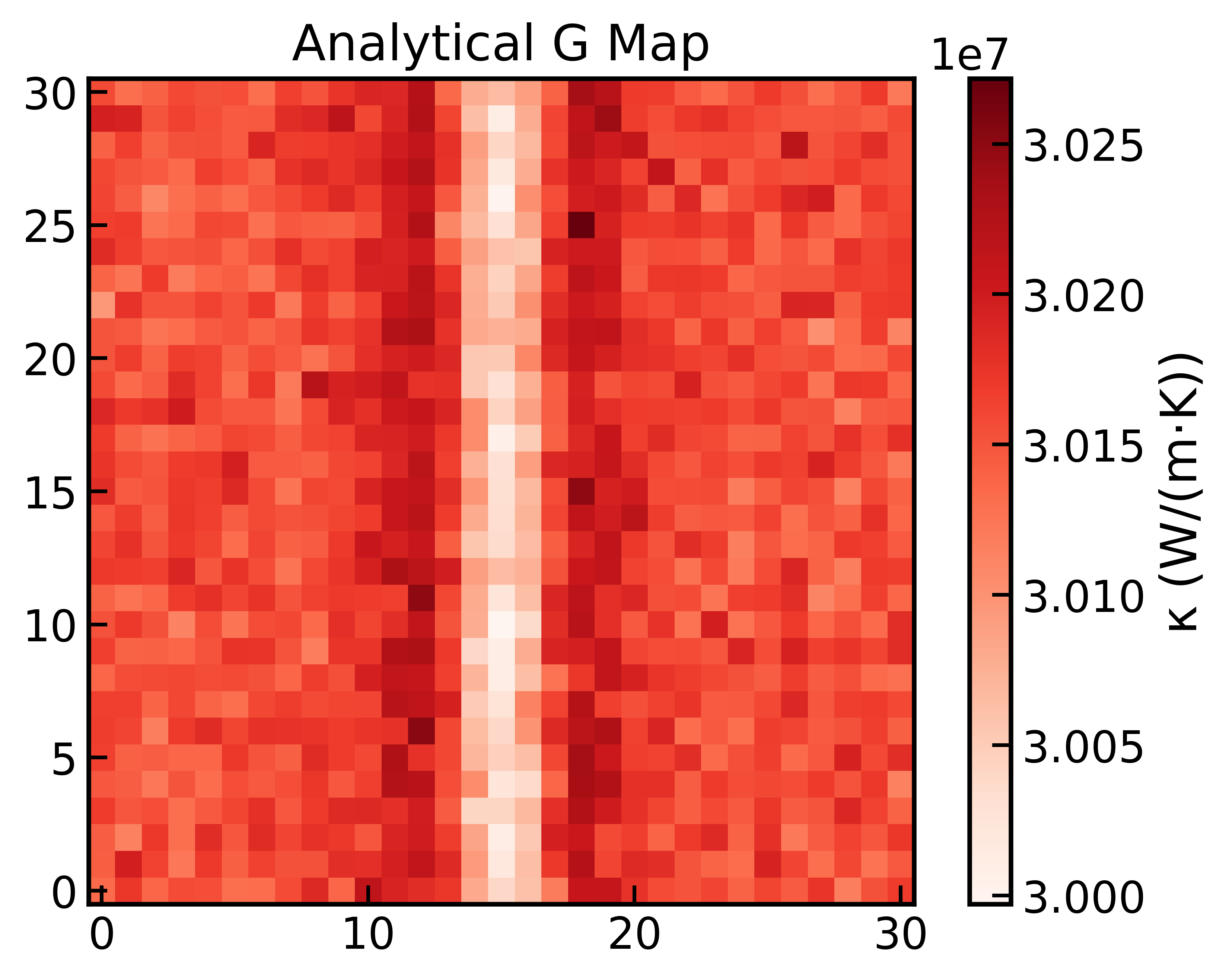} \\ 

        (a)  & 
        (b)  \\ 
        
    \end{tabular}
    \caption{Analytical $\kappa$ and $G$ fits}
    \label{fig:training_fit}
\end{figure}

Figure \ref{fig:training_fit} shows the fitted thermal conductivity and interface conductance obtained by minimizing the error between the predicted values and artificially generated experimental phase maps.

\begin{figure}[H]
\centering
\includegraphics[scale=0.4]{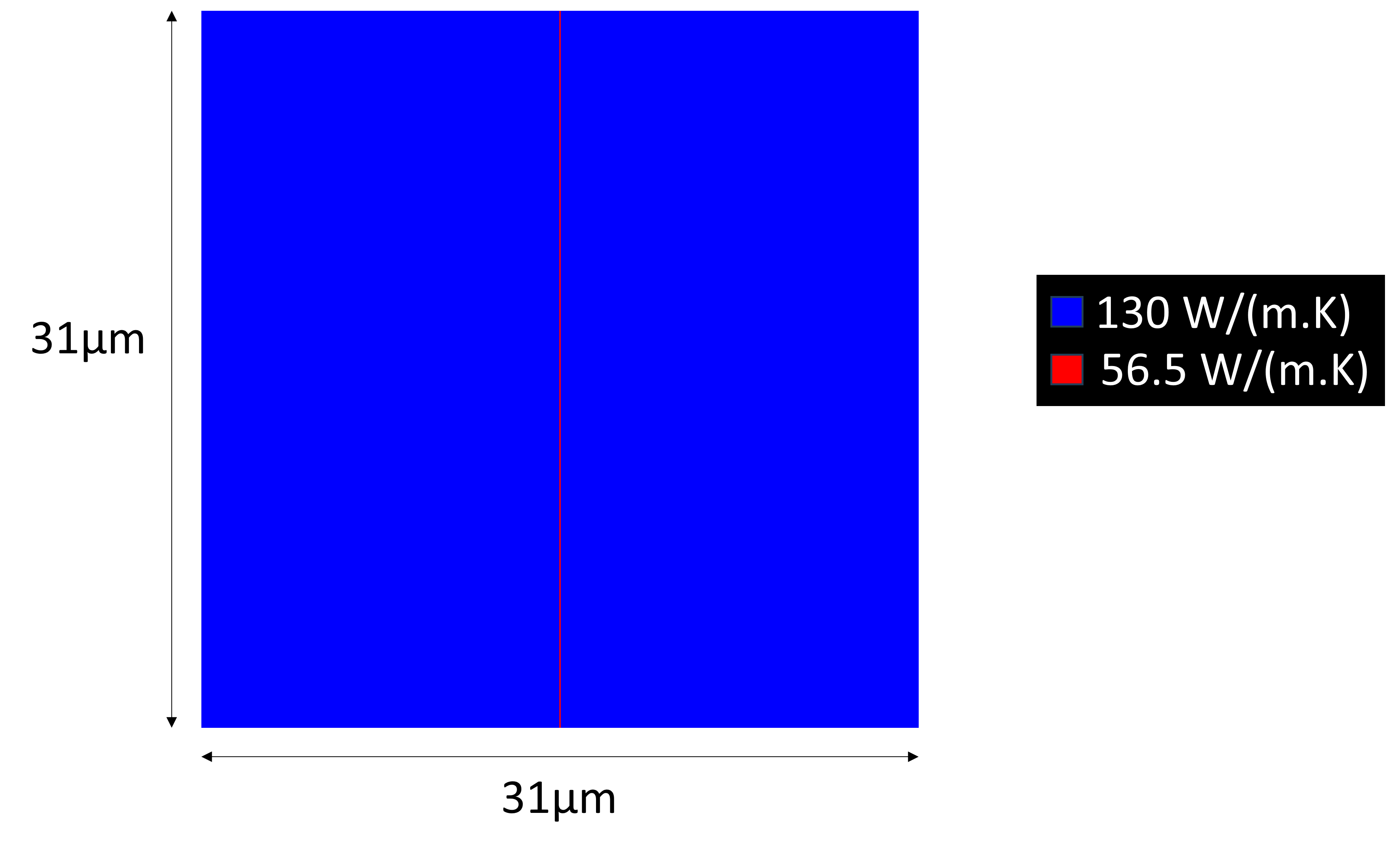}
\caption{FEM ``Ground Truth"}
\label{fig:training_ground_truth}
\end{figure}

\begin{figure}[H]
    \centering
    \begin{tabular}{ccc}
    
        \includegraphics[width=0.3\textwidth]{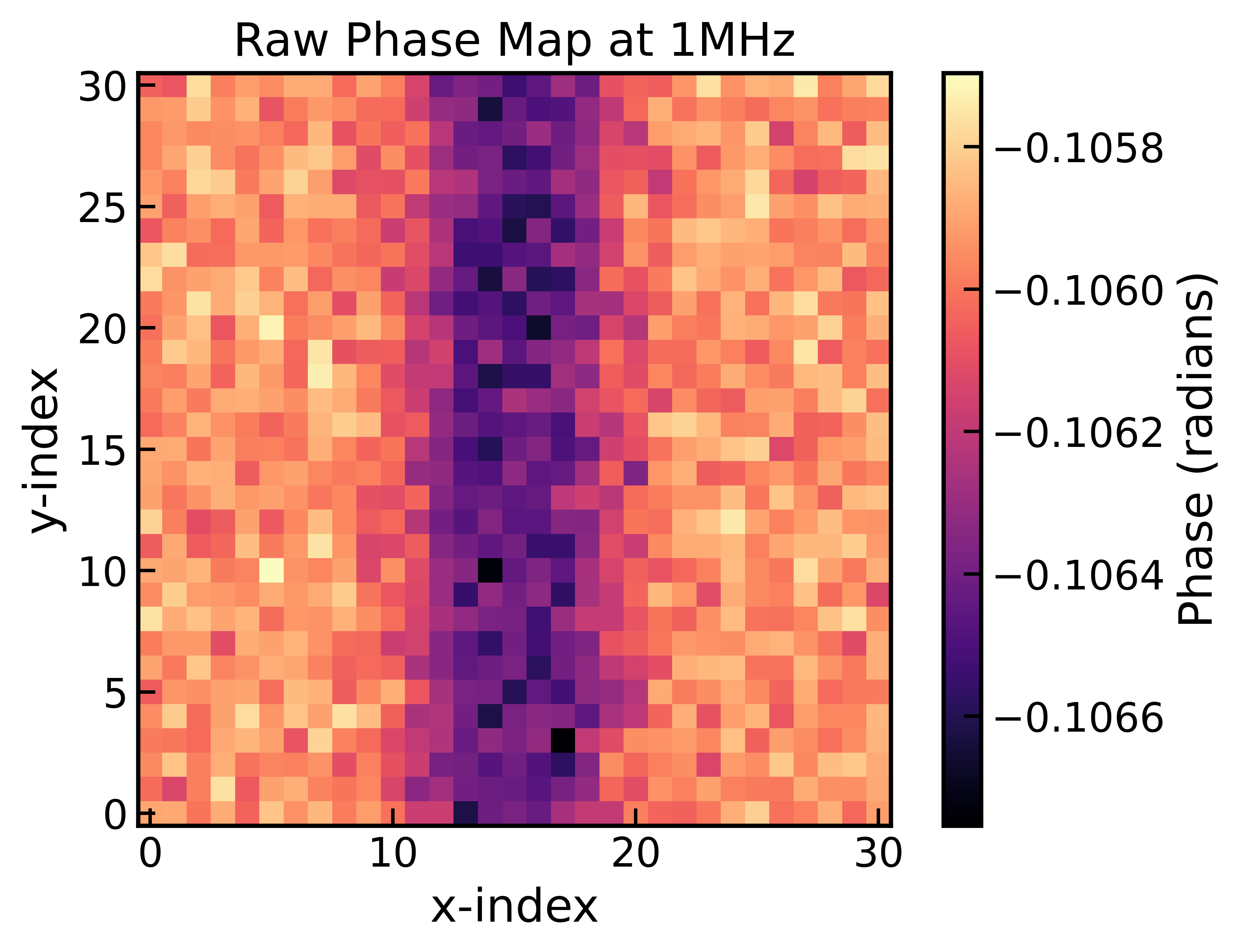} & 
        \includegraphics[width=0.3\textwidth]{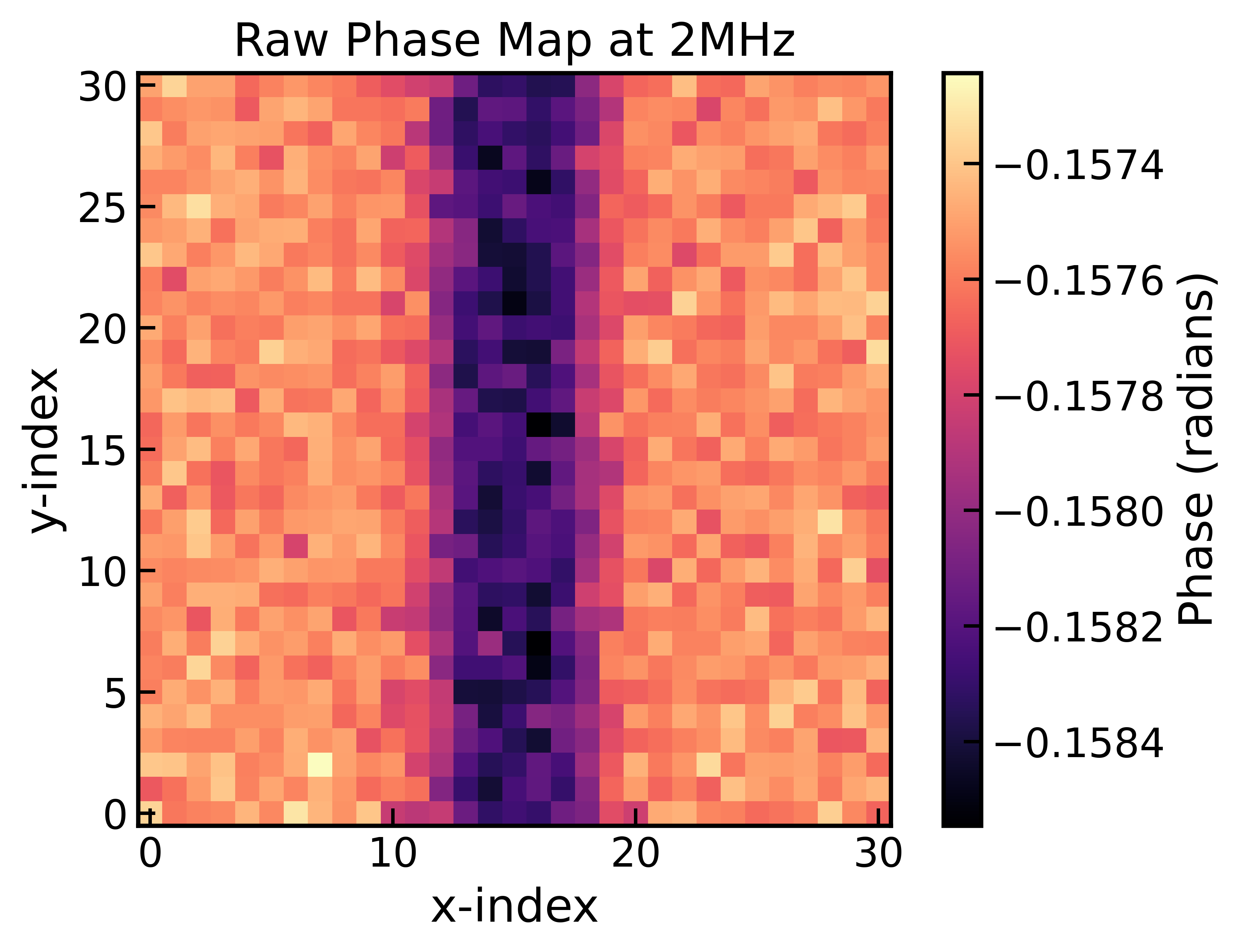} & 
        \includegraphics[width=0.3\textwidth]{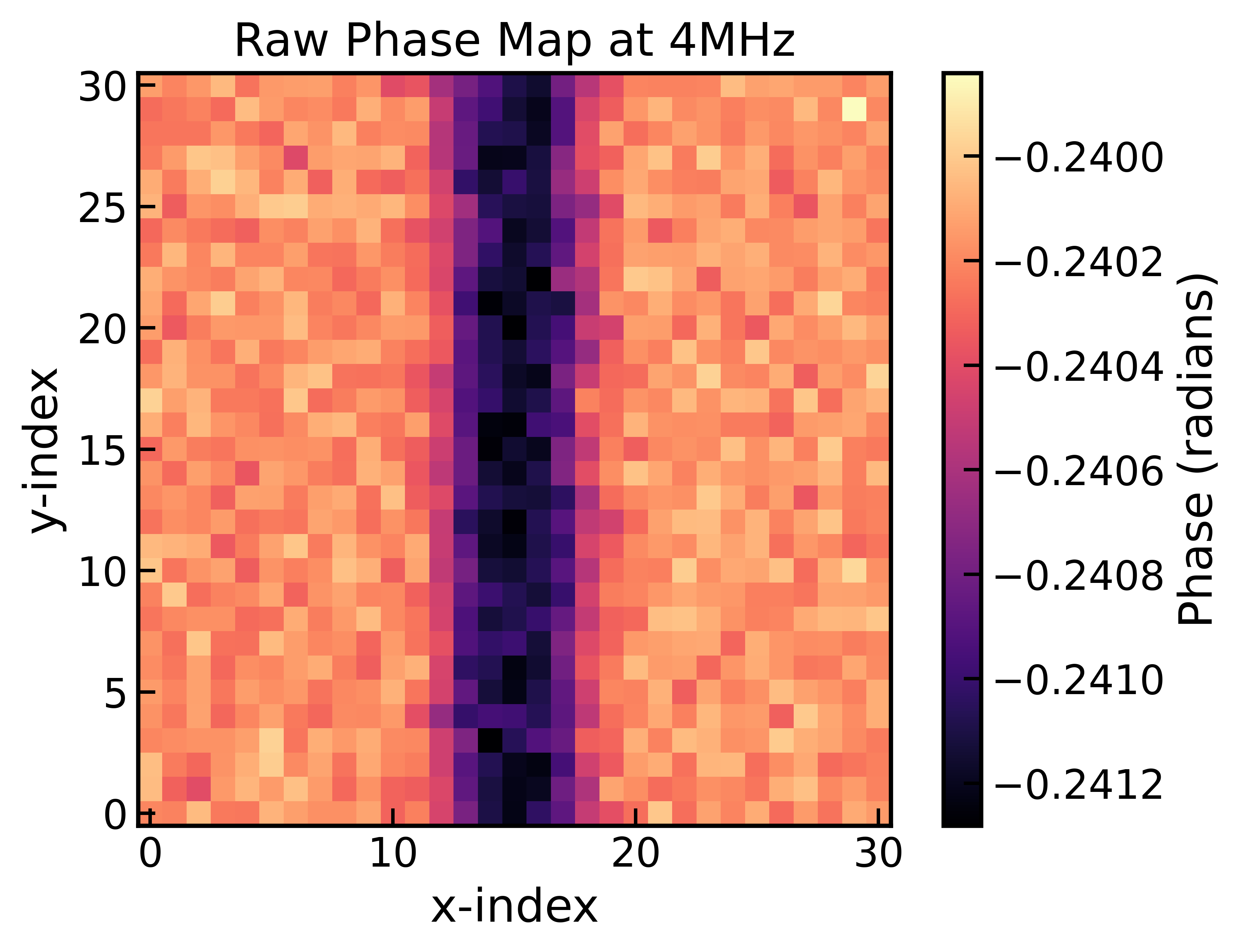} \\ 

        (a)  & 
        (b)  & 
        (c)  \\ 

        \includegraphics[width=0.3\textwidth]{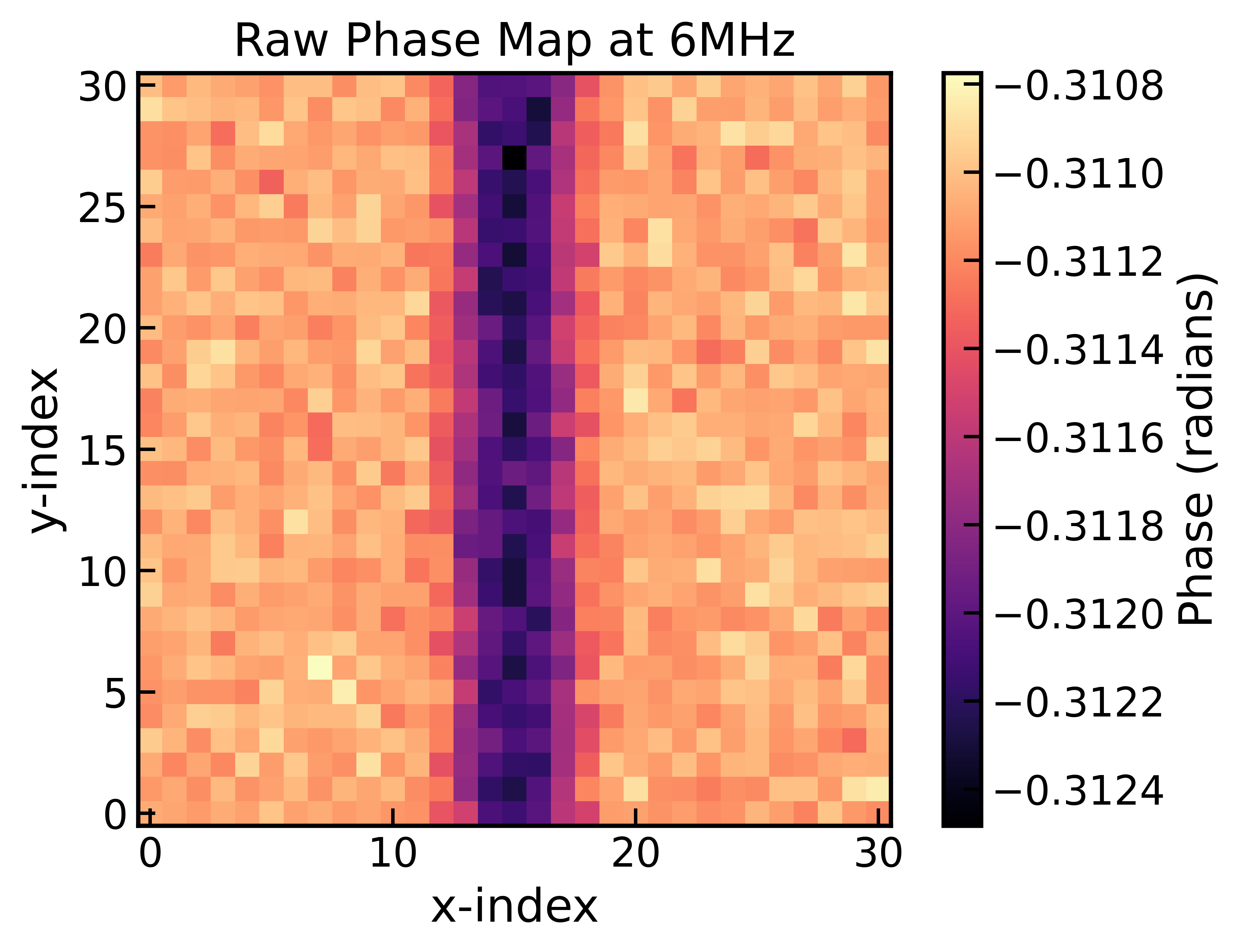} & 
        \includegraphics[width=0.3\textwidth]{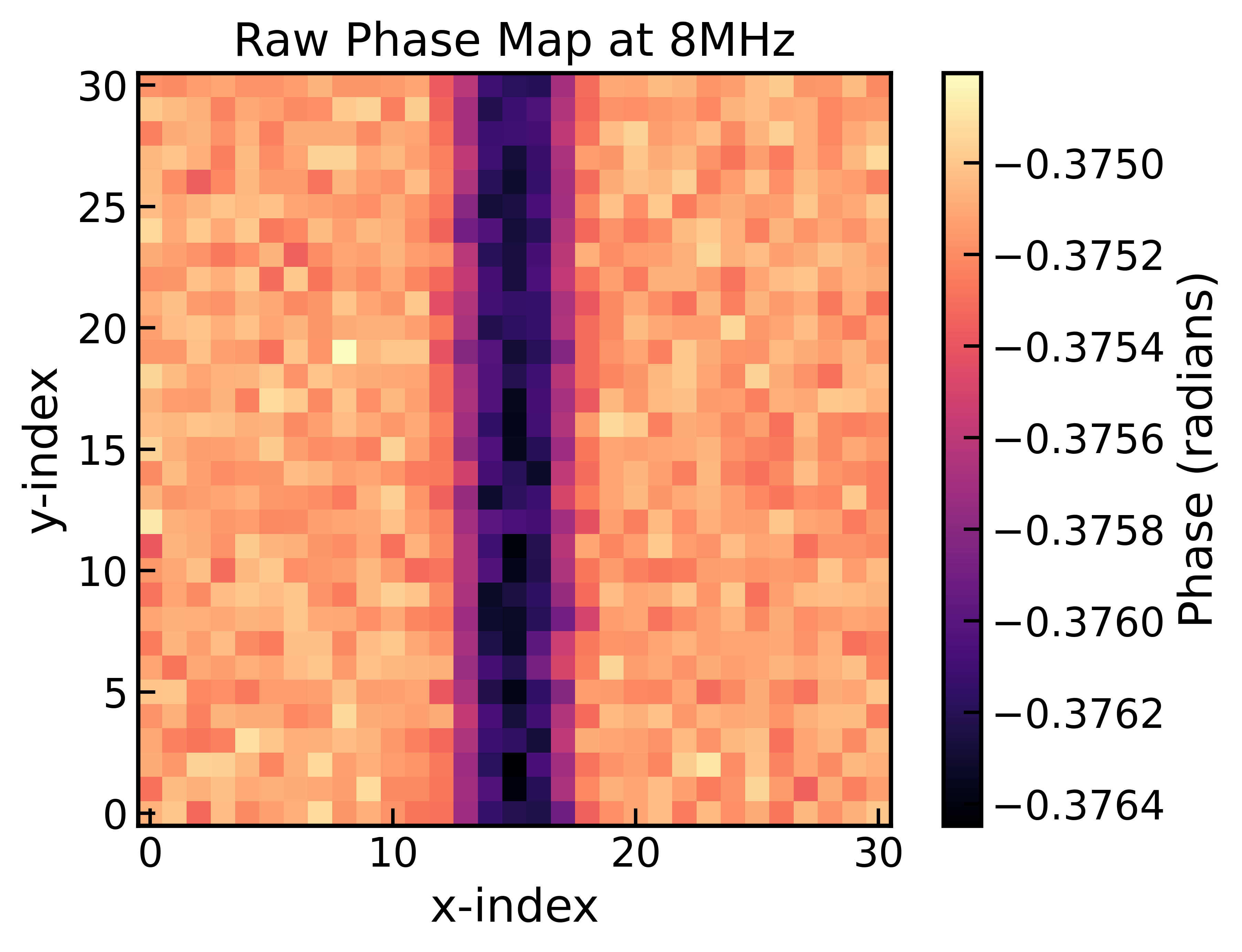} & 
        \includegraphics[width=0.3\textwidth]{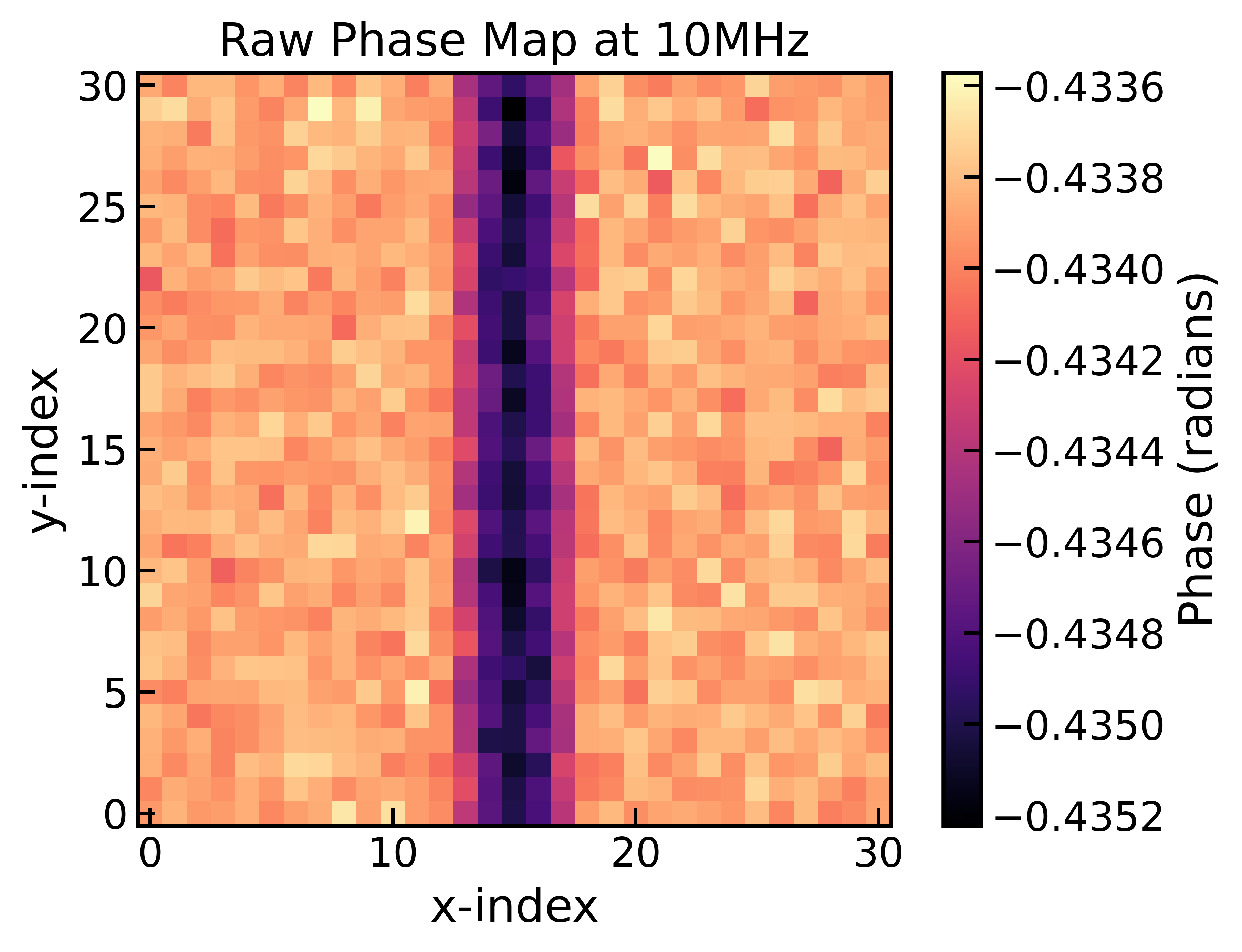} \\ 

        (d)  & 
        (e)  & 
        (f)  \\ 
        
    \end{tabular}
    \caption{FEM-generated ``experimental" phase maps}
    \label{fig:phasemaps}
\end{figure}

These ``experimental" phases were produced using finite element method (FEM) simulations with added random noise to mimic realistic measurement conditions, the details of which are included in the Appendix.  

It should be noted that the thermal conductivity map in Figure~\ref{fig:training_fit} was obtained by fitting each pixel independently. While this approach produces a map in which the grain-boundary pattern is visually discernible, it does not impose any spatial correlation between neighboring pixels. A physics-informed neural network (PINN) offers a natural extension to address this limitation - enabling the network to learn spatial patterns in the data, similar to the denoising approach demonstrated in Section~\ref{sec:1D_tutor}. A PINN incorporating the full FDTR physics would take the general loss form:

\begin{equation}
    \mathcal{L}_{\mathrm{PINN}} = 
    \underbrace{\sum (\phi^{\mathrm{pred}}(x,y,f) - \phi^{\mathrm{data}}(x,y,f))^2}_{\text{Data loss from neural network prediction}} +
    \underbrace{\sum (\phi^{\mathrm{pred}}(x,y,f) - \phi^{\mathrm{FDTR}}(x,y,f))^2}_{\text{Physics loss}}
    \label{eq:FDTR_pinn_loss_1}
\end{equation}

However, due to the discontinuities inherent in FDTR physics, the physics-loss term cannot be implemented directly as in the one-dimensional case. Instead, alternative formulations such as the Deep Energy Method (DEM)~\cite{samaniego2020} - potentially combined with domain-decomposition strategies used in fracture modeling~\cite{chakraborty2023} - may be employed to handle such discontinuous problems effectively.

For even greater accuracy, one could in principle compare the network-predicted phase directly with the analytical solution given in Equation~\ref{eq:phase}. That is,

\begin{equation}
    \mathcal{L}_{\mathrm{PINN}} = 
    \underbrace{\sum (\phi^{\mathrm{pred}}(x,y,f) - \phi^{\mathrm{data}}(x,y,f))^2}_{\text{Data loss from neural network prediction}} +
    \underbrace{\sum (\phi^{\mathrm{pred}}(x,y,f) - \phi^{\mathrm{analytical}}(x,y,f))^2}_{\text{Physics loss}}
    \label{eq:FDTR_pinn_loss_2}
\end{equation}

This was attempted but found to be prohibitively computationally expensive. To address this, we precompute the analytical solution and use the corresponding physics-based thermal conductivity map within a surrogate loss that compares the predicted and analytical thermal conductivities. Although this surrogate formulation omits direct differential operators, it remains physics-informed since the analytical inversion used to generate the reference thermal conductivity map embeds the full FDTR physics. The neural network thus learns to deconvolve a precomputed physics-consistent field rather than a purely data-driven one.

The following section details this surrogate-loss formulation, implemented through a Gaussian convolution that serves as a parameter-reduced representation of the full FDTR physics.

\subsection{FDTR Approximation - Gaussian Kernel}

As observed in the previous section, the thermal conductivity image obtained in Figure \ref{fig:training_fit} resembles a ``blurred" version of the ground truth presented in Figure \ref{fig:training_ground_truth}. This is similar to the ``blur'' obtained in continuum damage mechanics, where the damage/fracture due to strain localization at a sharp crack (the representation of which is limited by the size of the finite elements in continuum simulations) is regularized through a radial convolution of the strain, smearing out the damage.

\begin{figure}[H]
    \centering
    \begin{tabular}{ccc}
    
        \includegraphics[width=0.4\textwidth]{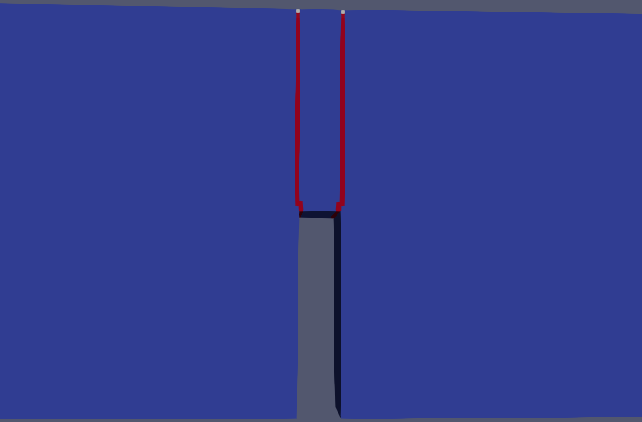} & 
        \includegraphics[width=0.4\textwidth]{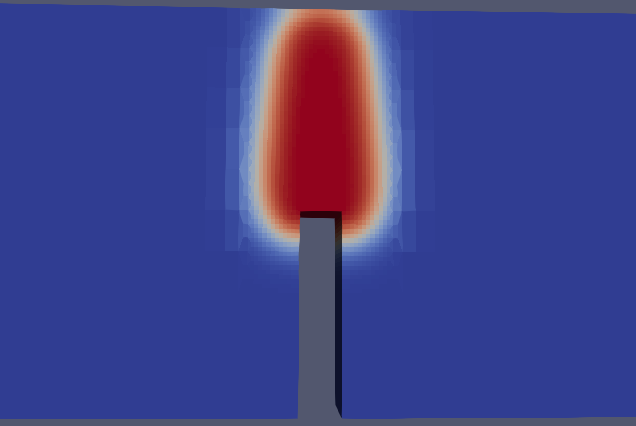} \\ 

        (a) No Strain Regularization & 
        (b) Smeared Damage \\ 
        
    \end{tabular}
    \caption{Comparison of Local and Nonlocal (Strain-Regularized) Continuum Damage Models}
    \label{fig:fracture}
\end{figure}

Therefore, because we intend to correlate high resolution structural images and blurry thermal conductivity maps, we can similarly abstract the FDTR physics as a convolution of the ground truth thermal conductivity with a multiplicative superposition of two Gaussian expressions, that is:

\begin{equation}
    \kappa^{FDTR} = \kappa^{struct} * \Psi
    \label{eq:convolution}
\end{equation}

\begin{figure}[H]
\centering
\includegraphics[scale=0.5]{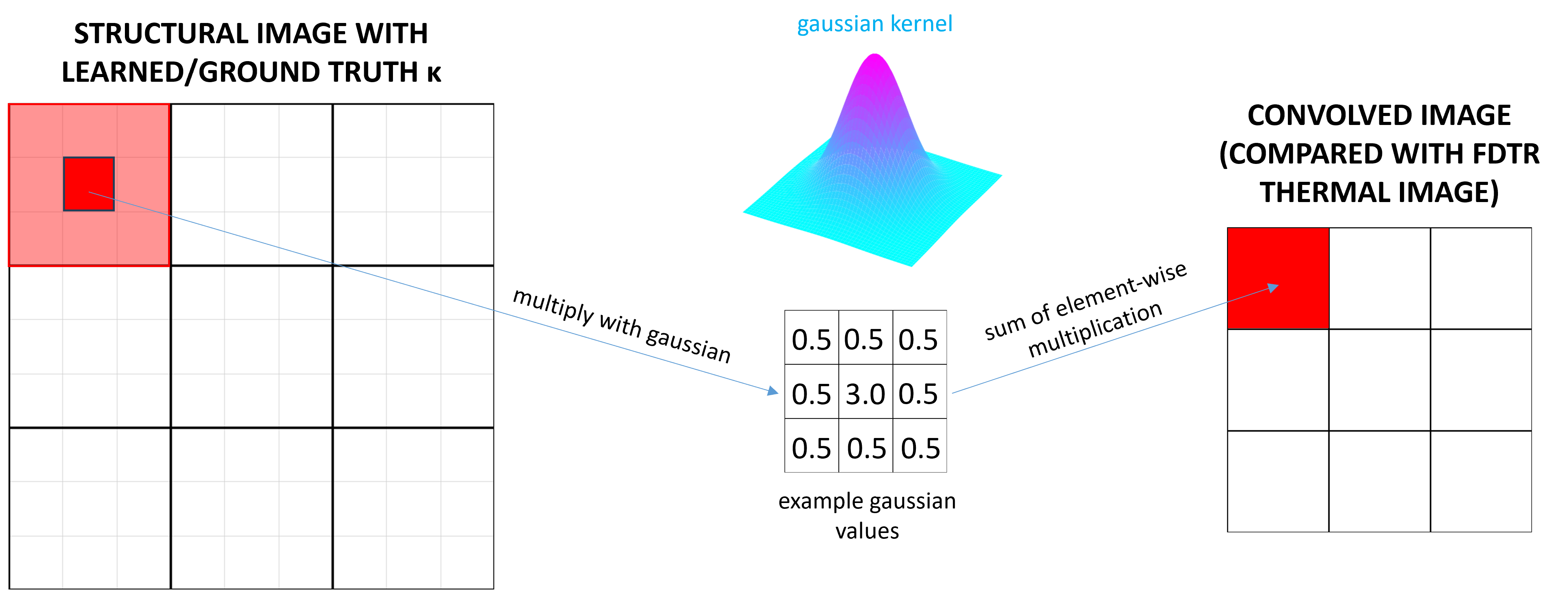}
\caption{Convolution of a high-resolution image to a lower-resolution image}
\label{fig:conv}
\end{figure}

With $\Psi$, the convolution ``kernel", defined as:

\begin{equation}
    \Psi = \exp{\left( \frac{-r^2}{2 \sigma_1} \right)} \exp{\left( \frac{-r^2}{2 \sigma_2} \right)}
    \label{eq:convolution_func}
\end{equation}

where $\sigma_1, \sigma_2$ are two parameters that describe the extent of the Gaussians and $r$ is the distance between the point being convolved and other points within the Gaussian support region. Note that the product of two Gaussian expressions is still a Gaussian. Figure \ref{fig:conv} clearly illustrates this.

In summary, we approximate the full FDTR physics as a two-parameter function defined as:

\begin{equation}
    f(\kappa^{struct}; \sigma_1, \sigma_2) \to \kappa^{FDTR}
    \label{eq:convolution_func_def}
\end{equation}

\begin{figure}[H]
    \centering
    \begin{tabular}{ccc}
    
        \includegraphics[width=0.3\textwidth]{images/Analytical_Kappa_Plot_Train.png} & 
        \includegraphics[width=0.3\textwidth]{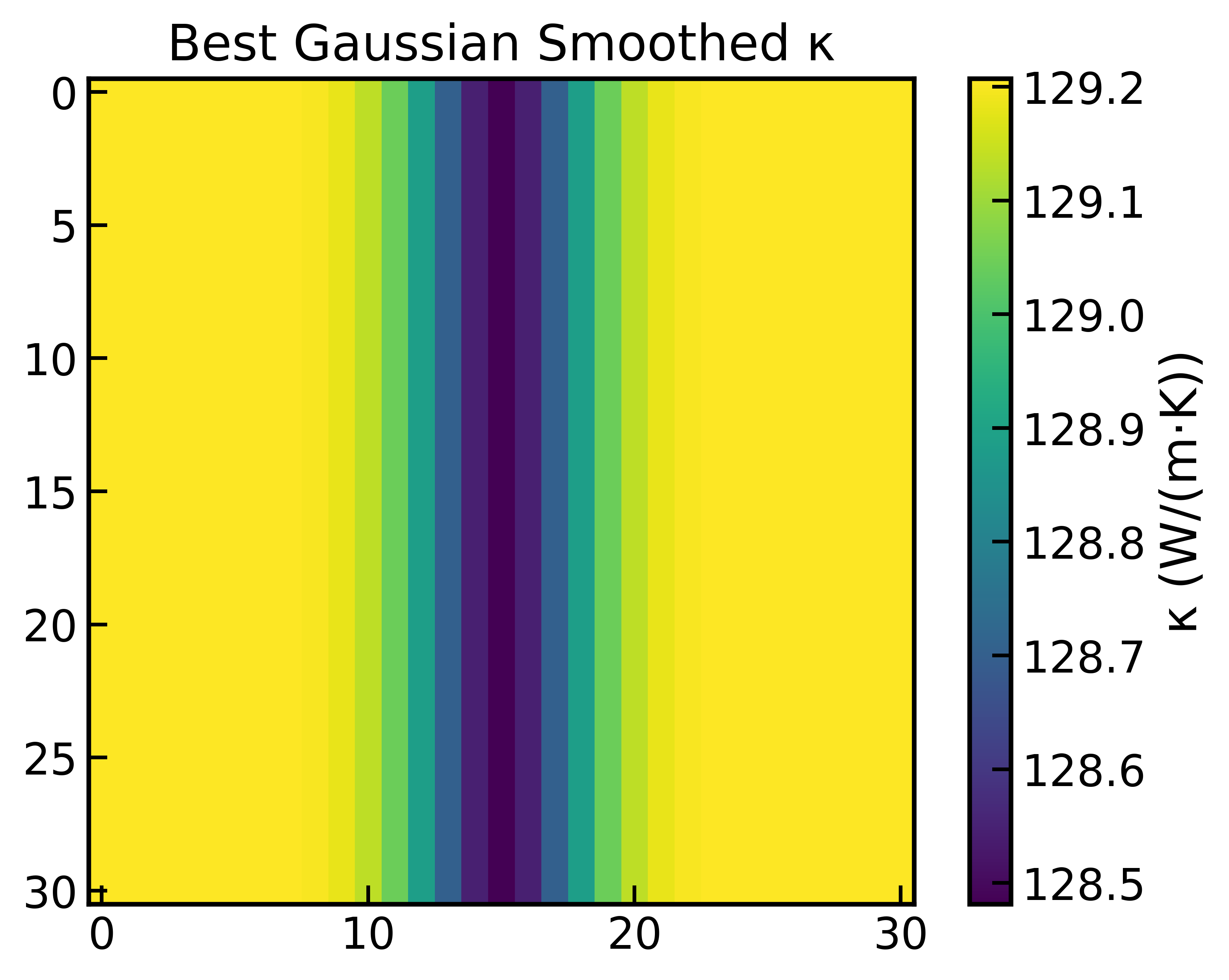} & 
        \includegraphics[width=0.3\textwidth]{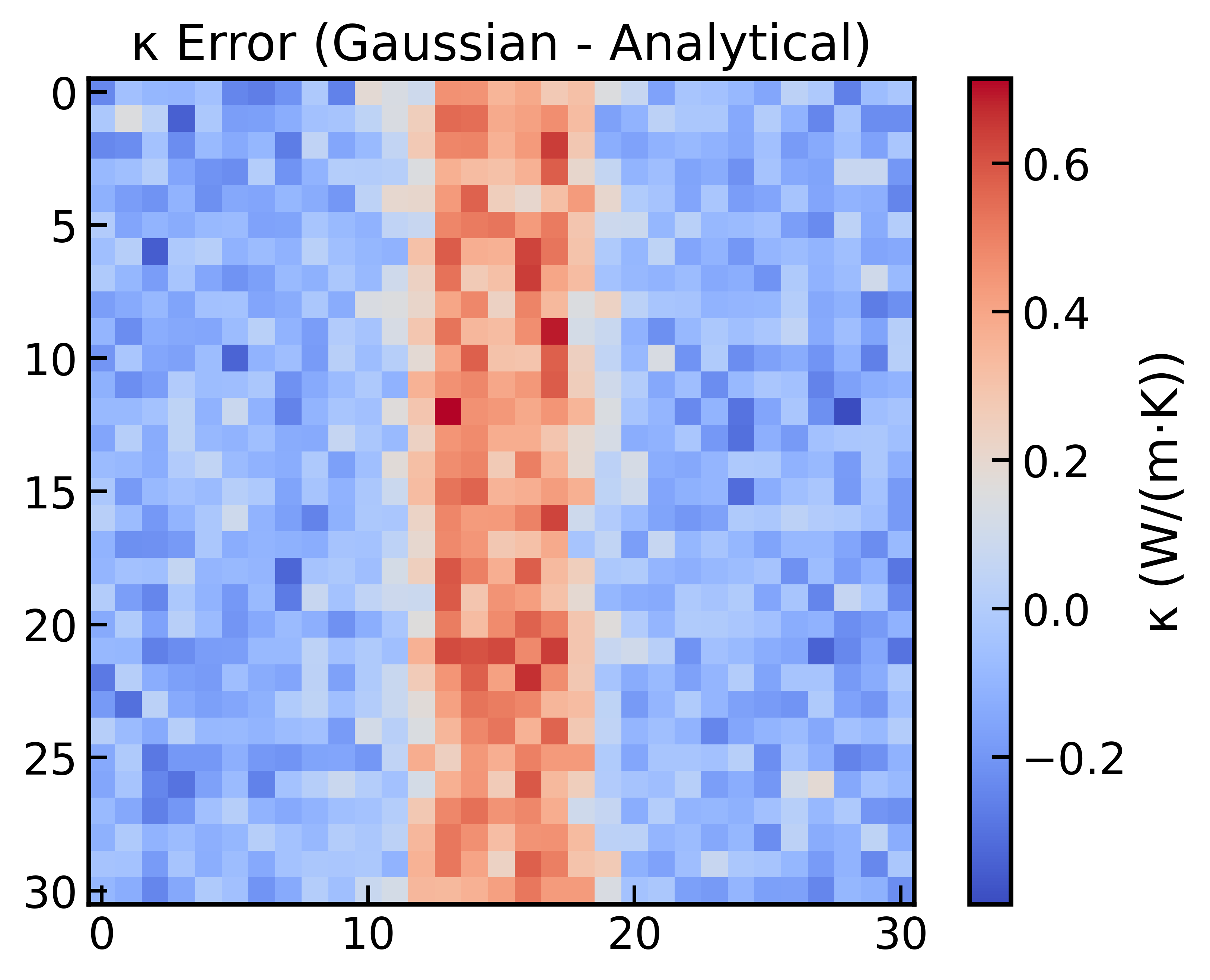} \\ 

        (a) Analytical $\kappa$ & 
        (b) Gaussian Smoothed $\kappa$ & 
        (c) Error \\ 
        
    \end{tabular}
    \caption{``Denoised" thermal image through Gaussian convolution based on known ground truth. (RMSE: $0.22$ ~W/(m$\cdot$K))}
    \label{fig:gaussian_smooth}
\end{figure}

There is a mathematical justification for this approximation detailed in the Appendix. Applying this to synthetic FDTR data with a known ground truth (Figure \ref{fig:training_ground_truth}), Figure \ref{fig:gaussian_smooth} shows an inverse fit to obtain $\sigma_1$ and $\sigma_2$, demonstrating the efficacy of this approximation.

For this case, the fitted values were $\sigma_1 = 113.4995$ and $\sigma_2 = 104.9994$, obtained using the Nelder--Mead algorithm, which was selected for its robustness in optimizing noisy, non-differentiable functions. However, repeated attempts with alternative optimization methods revealed that the inverse problem is underconstrained - multiple combinations of $\sigma_1$ and $\sigma_2$ yield nearly identical Gaussian-smoothed profiles. This underconstraint arises because multiple Gaussian parameter pairs can produce nearly identical smoothing effects, particularly when the underlying image resolution limits the ability to distinguish fine-scale variations. Thus, while the Gaussian abstraction provides a convenient reduced-order model, it should be interpreted as a physically informed regularization rather than a unique inversion.

This finding reinforces that the present approach should be viewed as a proof-of-concept for abstracting FDTR physics. Future implementations may employ supervised convolutional neural networks (CNNs), using this Gaussian approximation as an effective initial condition, to more accurately capture the full FDTR physics and enable rapid ``virtual'' FDTR experiments once thoroughly validated.

Figure \ref{fig:gaussian_smooth} shows smoothed $\kappa$ profiles for different $\sigma$ values obtained with \texttt{GaussianSmoother.py}.

\begin{figure}[H]
    \centering
    \begin{tabular}{ccc}
    
        \includegraphics[width=0.3\textwidth]{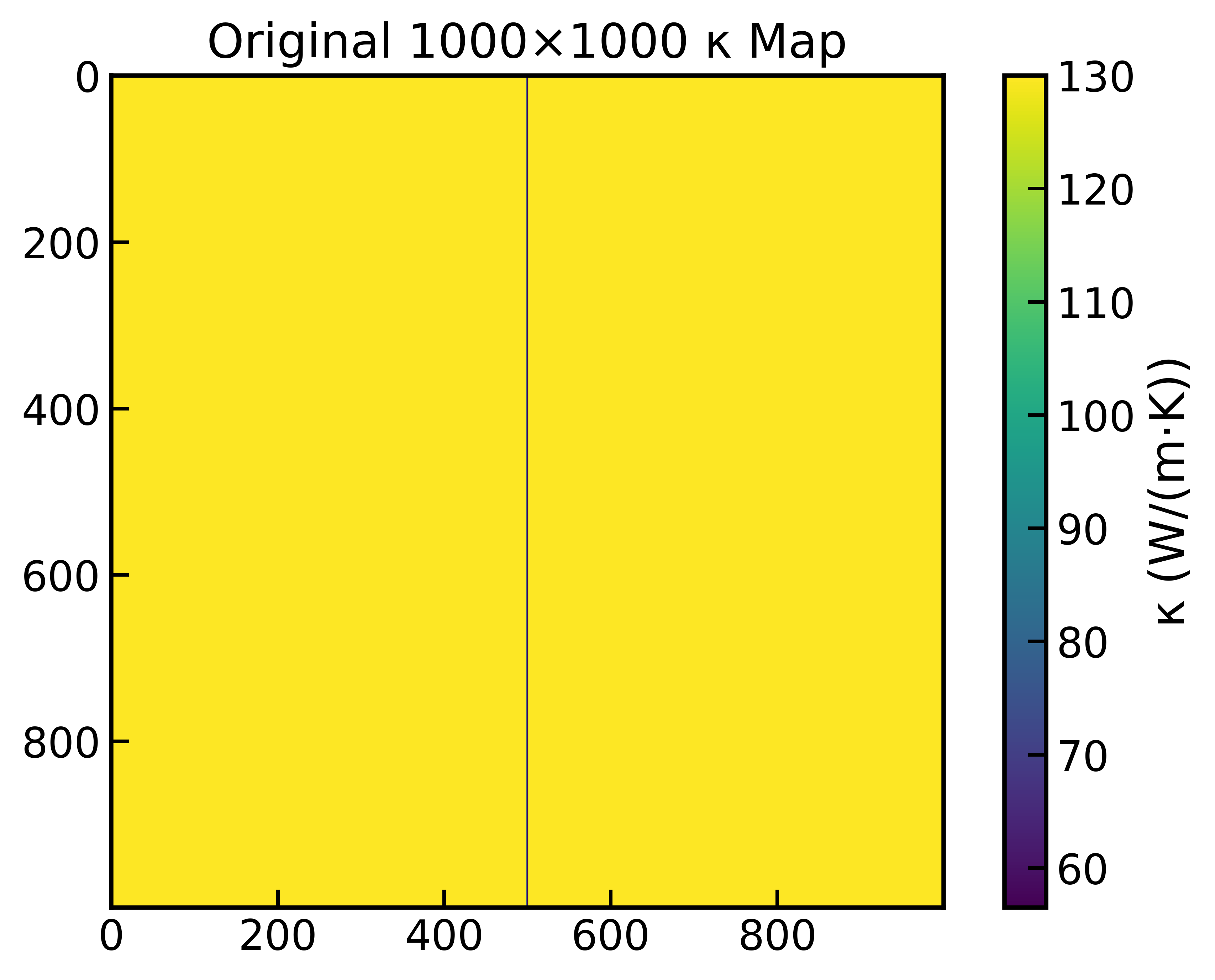} & 
        \includegraphics[width=0.3\textwidth]{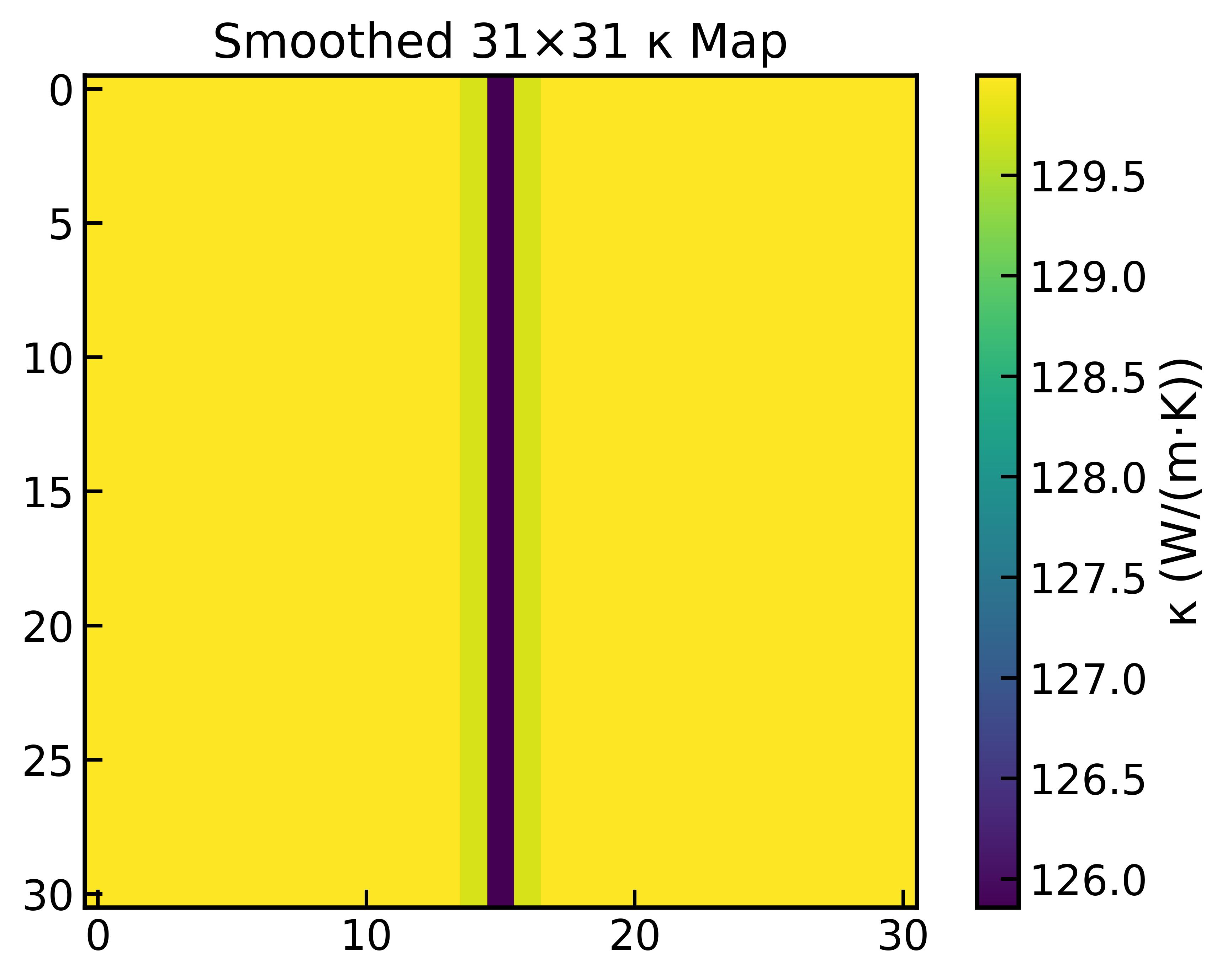} & 
        \includegraphics[width=0.3\textwidth]{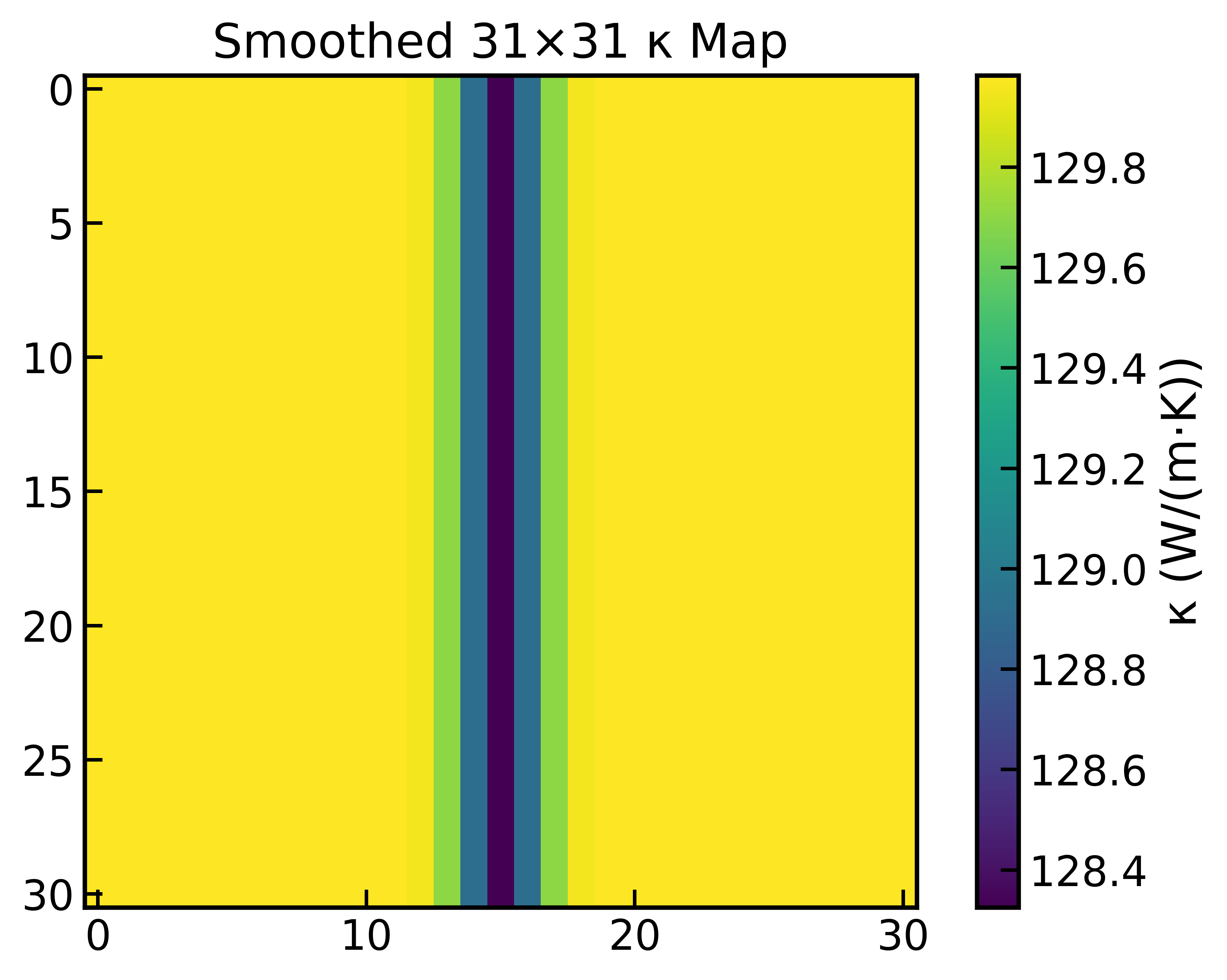} \\ 

        (a) Original Structure & 
        (b) $\sigma_1 = \sigma_2 = 20$ & 
        (c) $\sigma_1 = \sigma_2 = 50$ \\ 
        
    \end{tabular}
    \caption{Gaussian smoothing for different $\sigma$ values}
    \label{fig:gaussian_smooth_change_sigma}
\end{figure}

\section{Inverse Modeling With Gaussian Constrained Neural Network}

Now that we have abstracted the FDTR physics, we can now create a parameter-constrained neural network such that:

\begin{equation}
    \mathcal{L}_{PINN}(\kappa^{struct}, G) = \underbrace{ \sum (\phi^{pred}(x,y,f) - \phi^{data}(x,y,f))^2}_{\text{Data Loss from Neural Network Prediction}} + \underbrace{\sum ([\kappa^{struct} * \Psi] - \kappa^{analytical})^2}_{\text{``Physics" (Parameter) Loss}}
    \label{eq:FDTR_pinn_loss}
\end{equation}

Recall that $\kappa^{struct}$ is a high resolution matrix of $\kappa$ values which depends on the resolution of the structural image. Because this results in a highly underdetermined system, we significantly reduced the number of parameters to be learned by using k-means clustering to group the structural image into regions. This reduced-parameter approach preserves the essential microstructural variation while keeping the optimization problem computationally tractable. With the Gaussian convolution encapsulating FDTR’s spatial averaging and the microstructure-aware parameterization constraining $\kappa^{struct}$, the network is now ready for evaluation. 

In the next section, we apply this model to synthetic FDTR datasets generated from finite element simulations, including a more challenging heterointerface configuration, to assess its ability to recover bulk and interfacial thermal conductivities.

\section{Results and Discussion}

\begin{figure}[H]
\centering
\includegraphics[scale=0.4]{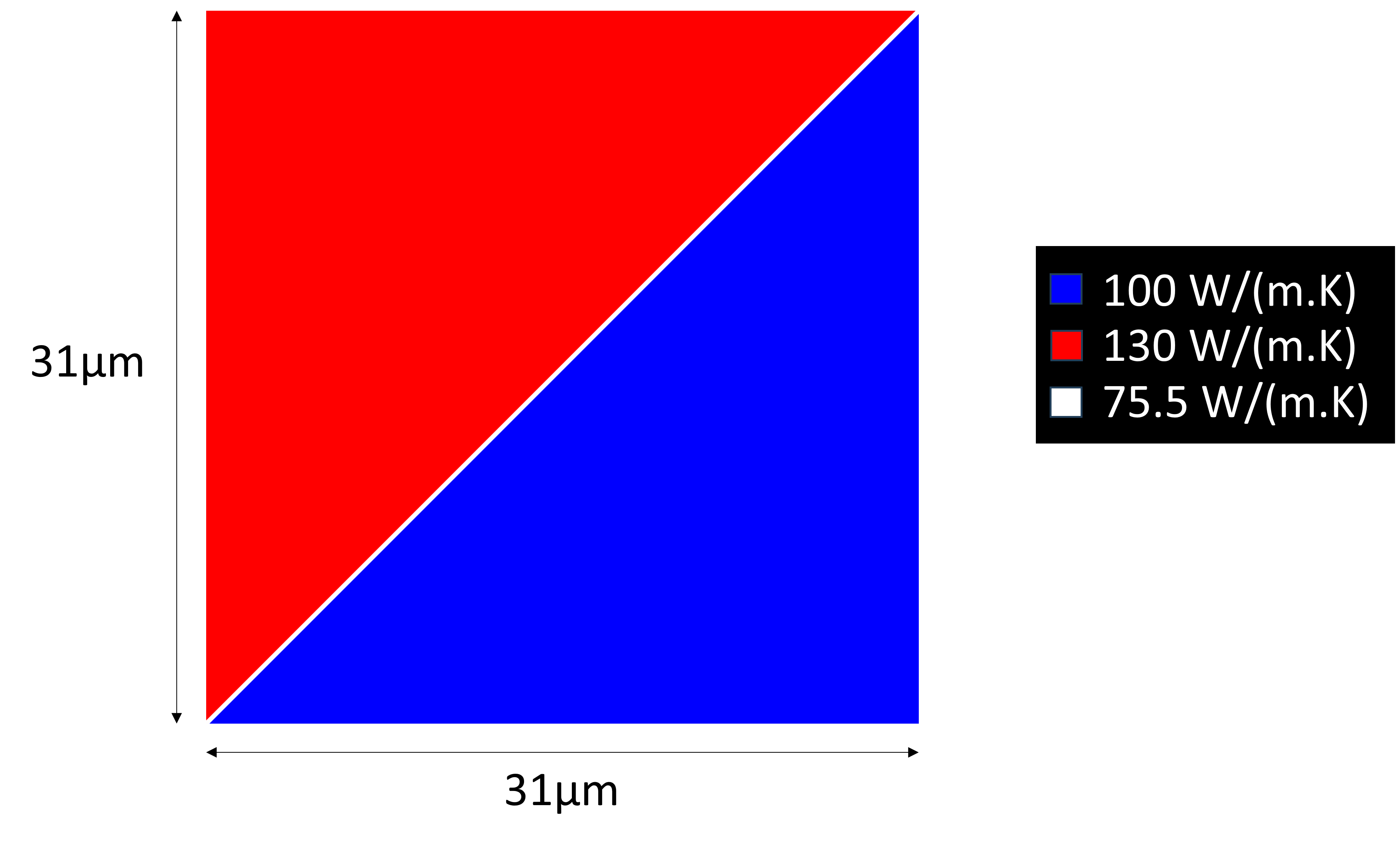}
\caption{FEM ``Ground Truth"}
\label{fig:test_ground_truth}
\end{figure}

\begin{figure}[H]
    \centering
    \begin{tabular}{ccc}
    
        \includegraphics[width=0.45\textwidth]{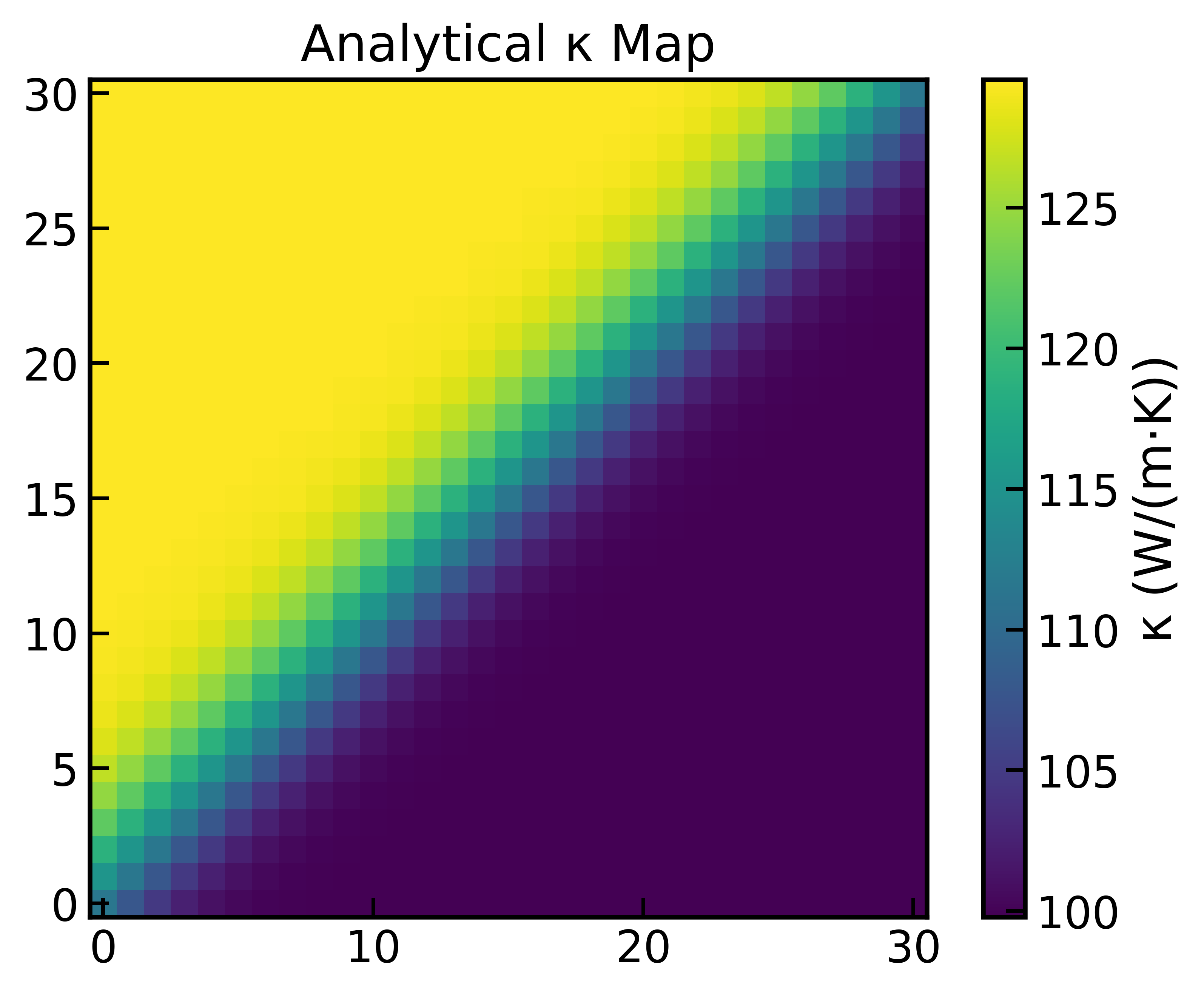} & 
        \includegraphics[width=0.45\textwidth]{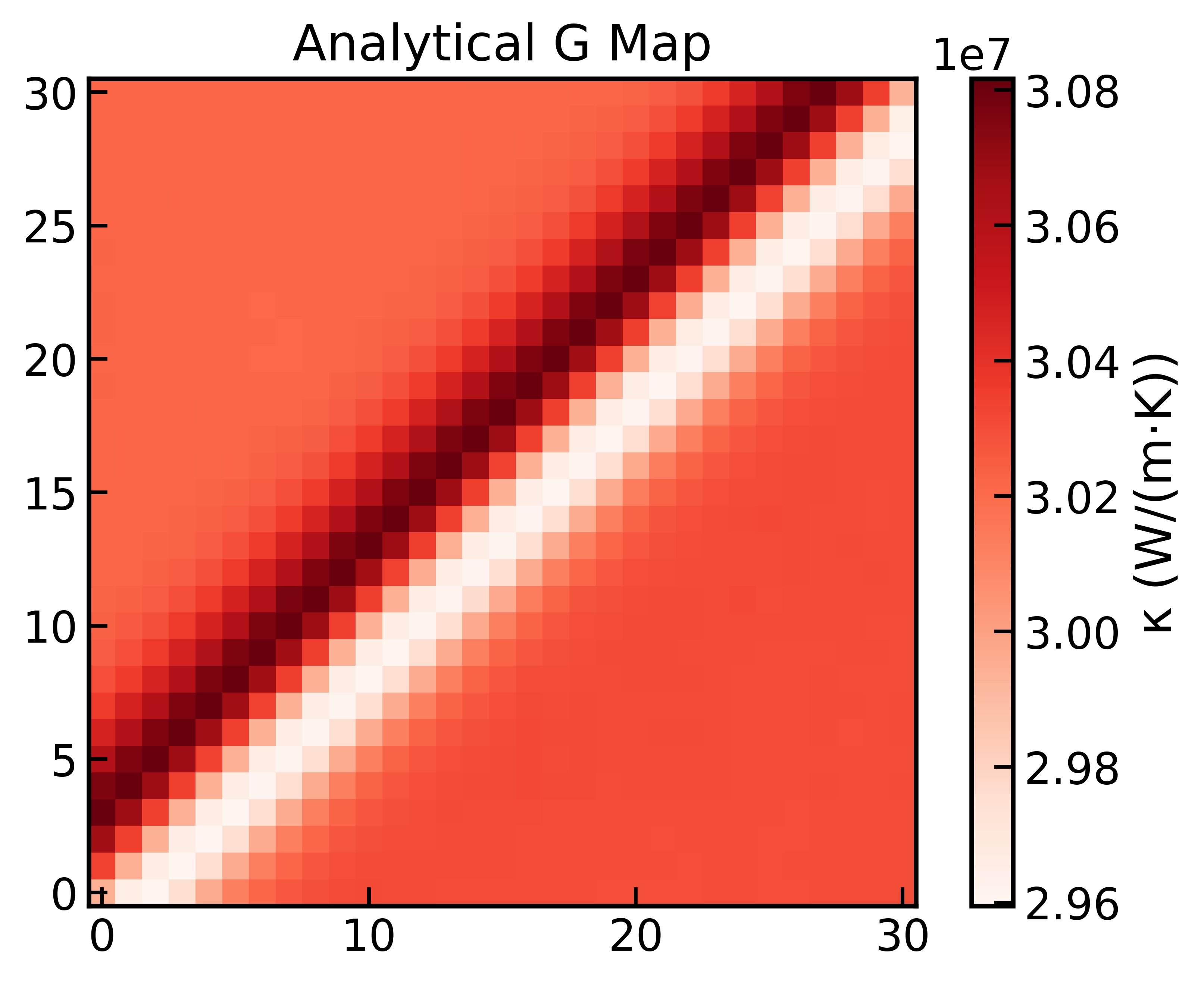} \\ 

        (a)  & 
        (b) \\ 
        
    \end{tabular}
    \caption{Analytical $\kappa$ and $G$ fits on test data}
    \label{fig:test_fit}
\end{figure}

To evaluate the performance of the framework, we applied it to synthetic FDTR datasets generated from finite element method (FEM) simulations, with the FEM implementation details laid out in the Appendix. The test case features a heterointerface - two materials with different thermal conductivities separated by a grain boundary - chosen to probe the ability of the model to recover both bulk and interfacial properties under more complex conditions. Figure~\ref{fig:test_ground_truth} shows the structural ground truth, whereas Figure~\ref{fig:test_fit} presents the corresponding analytically derived thermal conductivity map.

A k-means clustering algorithm was applied in RGB color space to identify a limited set of distinct color regions in the image. This reduced the number of unique thermal conductivities to be solved for, since each color cluster was treated as a region with a single effective conductivity.

Through the k-means clustering operation performed on the structural image, the model had only three distinct values of $\kappa$ to fit. Interestingly, although a grain boundary was simulated in between both regions, the resolution of the FDTR thermal image obscured this boundary. That is, without structural information, the thermal conductivity profile appears to be a smooth transition between $130$ and $100$ ~W/(m$\cdot$K) as shown in Figure \ref{fig:test_fit}.

\begin{figure}[H]
\centering
\includegraphics[scale=0.5]{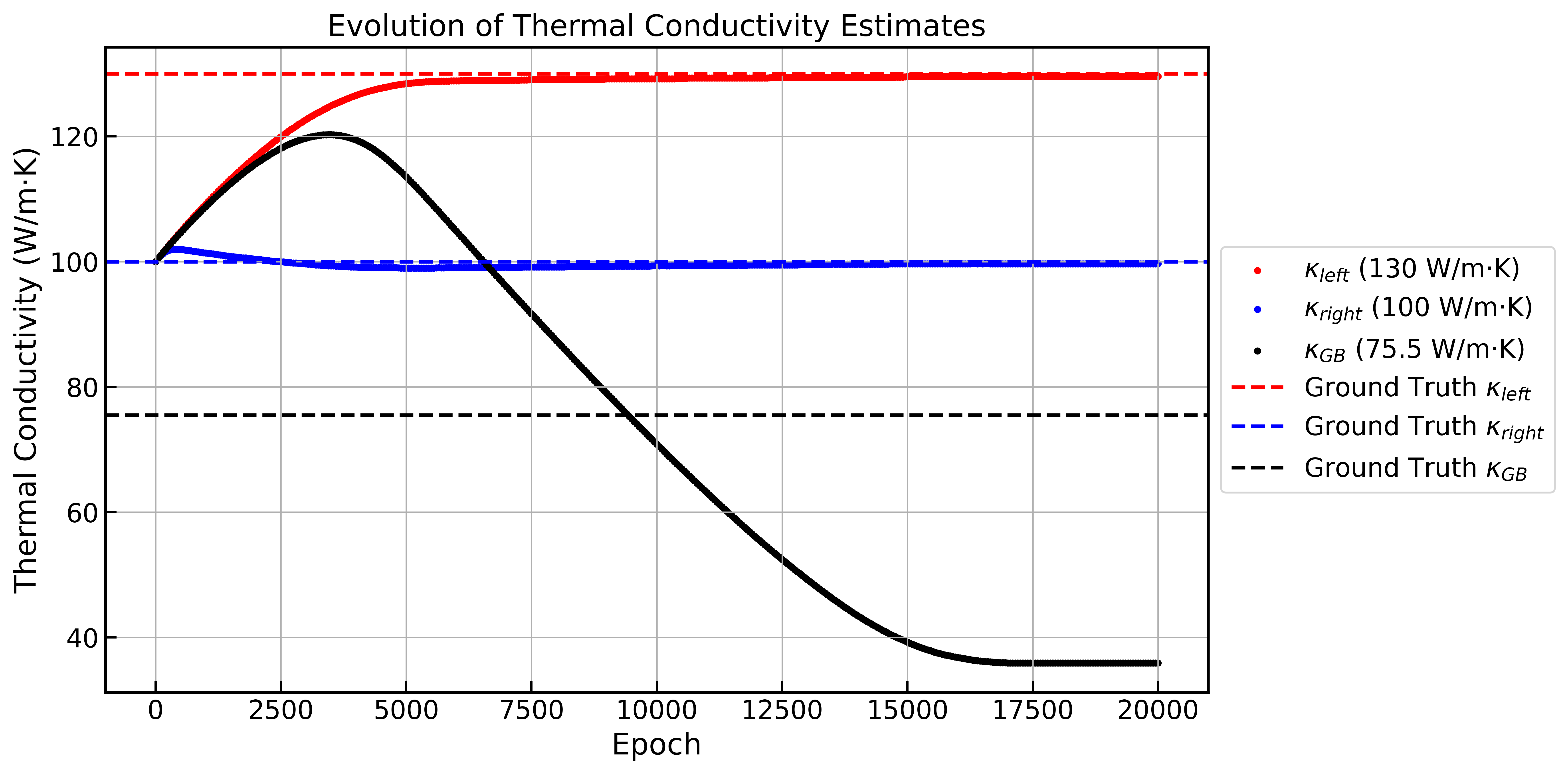}
\caption{Neural Network Predicted Thermal Conductivity vs Epoch}
\label{fig:kappa_evolve}
\end{figure}

\begin{figure}[H]
    \centering
    \begin{tabular}{ccc}
    
        \includegraphics[width=0.4\textwidth]{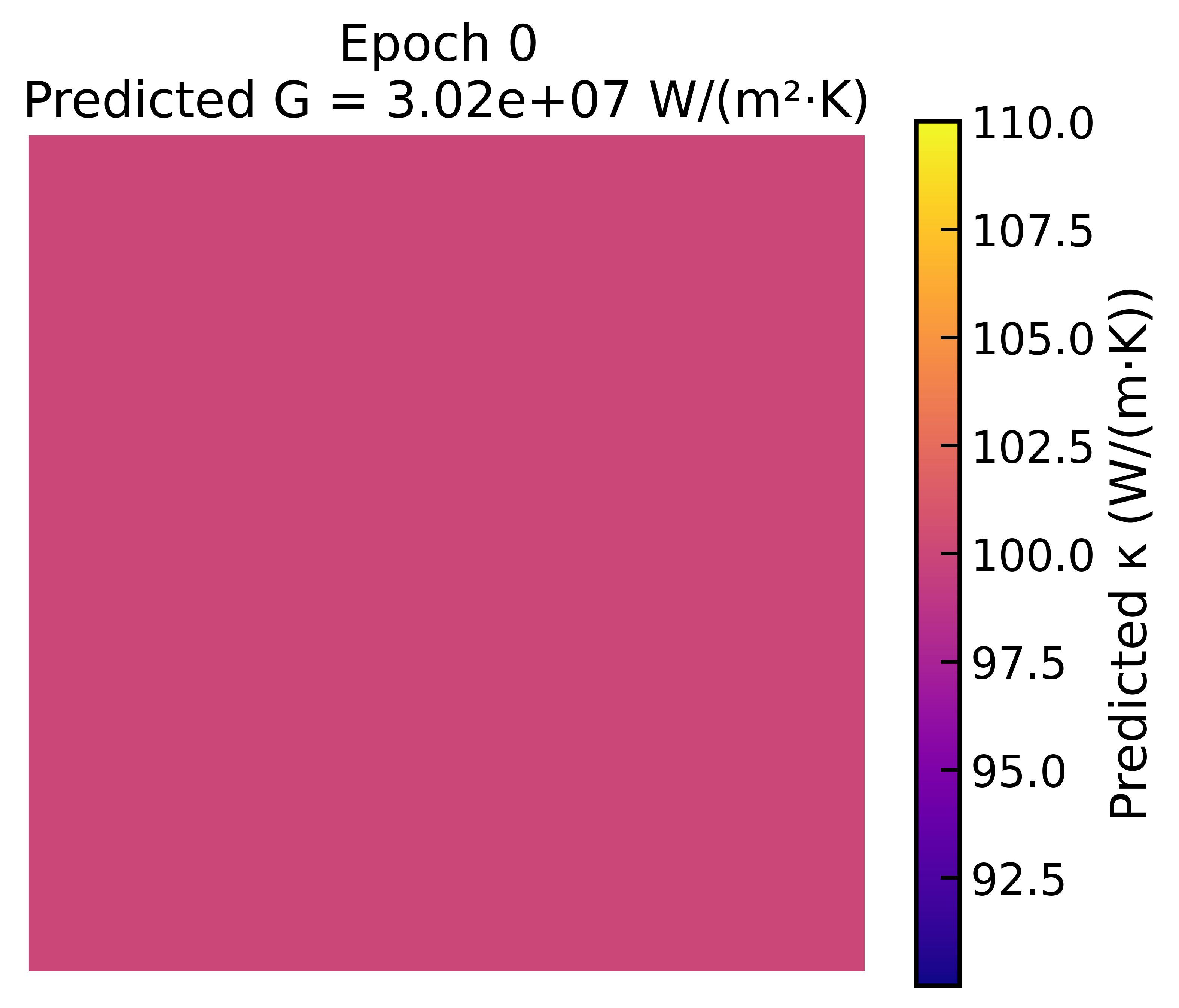} & 
        \includegraphics[width=0.4\textwidth]{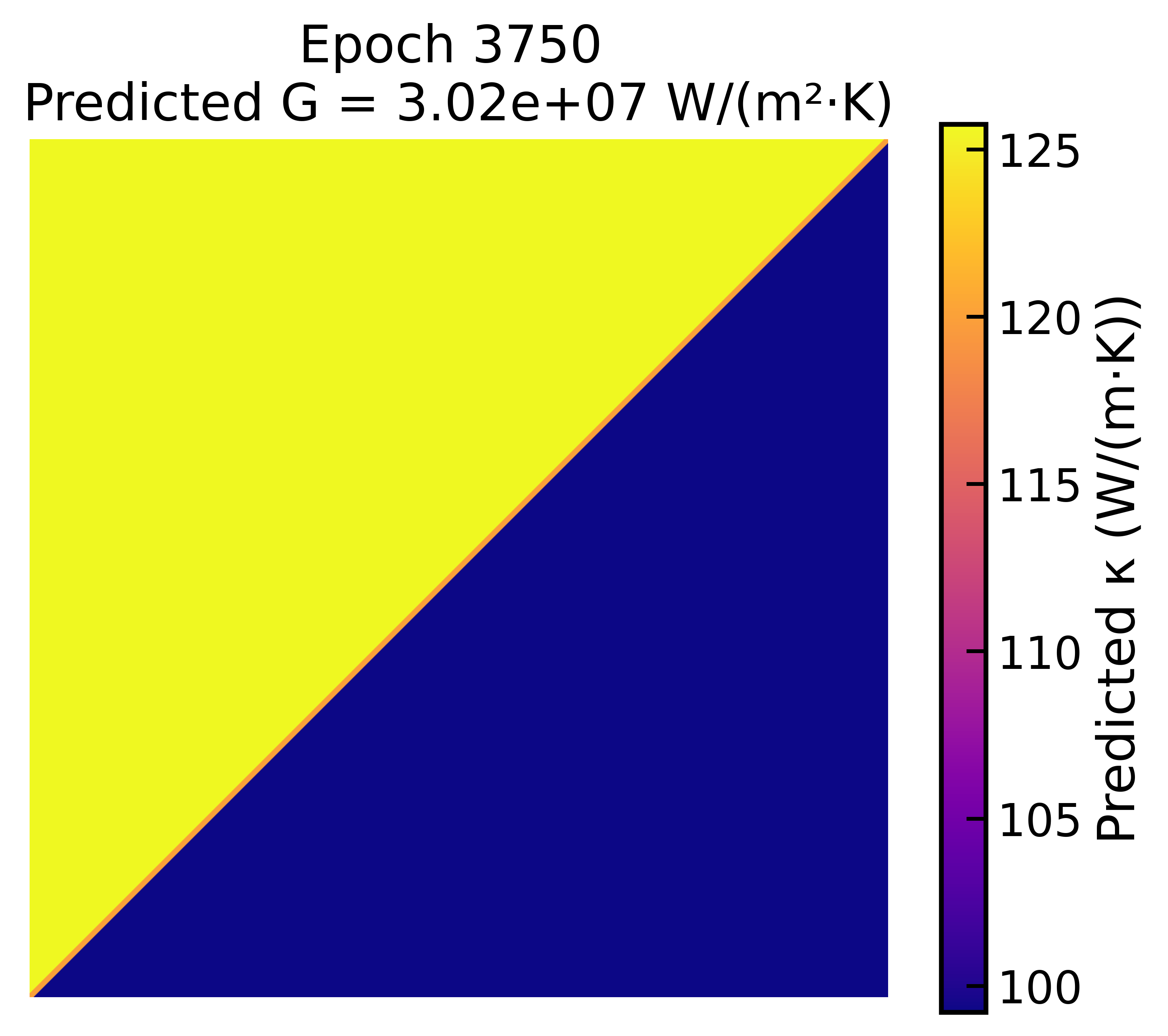} \\ 

        (a) $\kappa$ at Epoch $0$ & 
        (b) $\kappa$ at Epoch $3,750$ \\
        
        \includegraphics[width=0.4\textwidth]{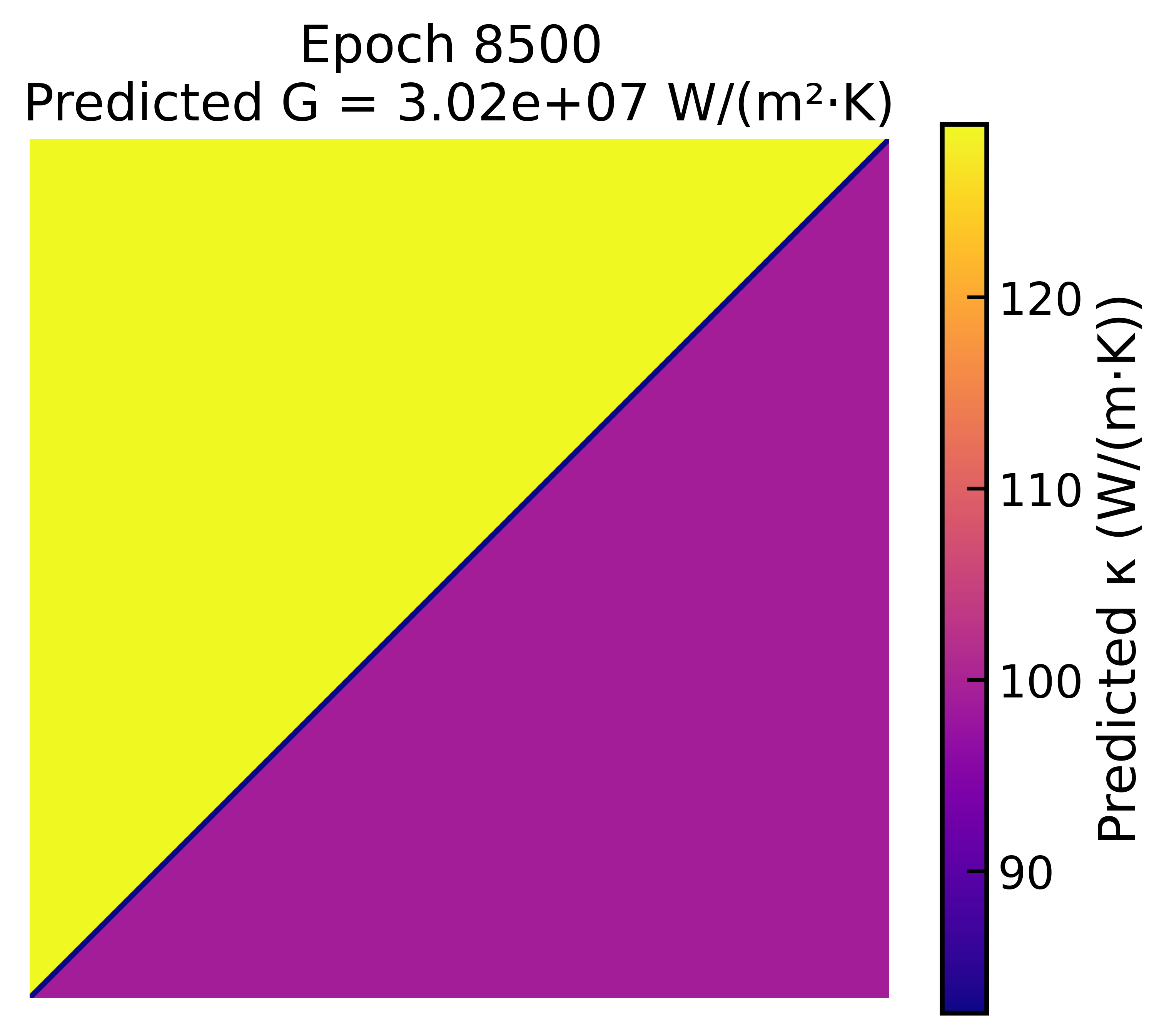} &
        \includegraphics[width=0.4\textwidth]{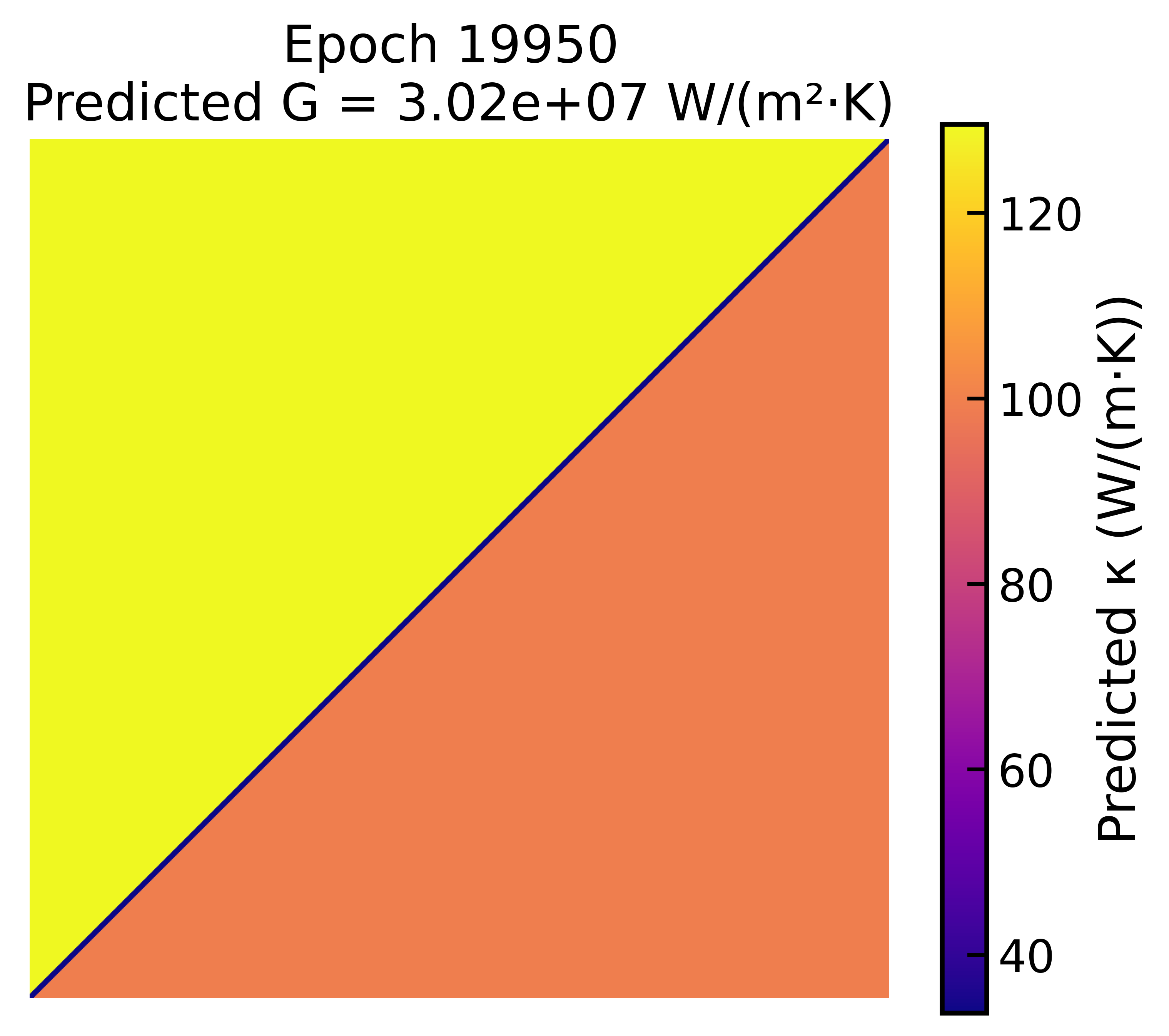} \\
        
        (a) $\kappa$ at Epoch $8,500$ & 
        (b) $\kappa$ at Epoch $19,950$ \\ 
        
    \end{tabular}
    \caption{Recovered Thermal Conductivity Profile Evolution}
    \label{fig:epoch_evolve}
\end{figure}

The training was terminated after $20,000$ epochs (training loops), once all three thermal conductivities ($\kappa$ values) had converged. The process required only a few minutes of runtime on an NVIDIA A100 GPU (via Google Colab Pro+), owing to the optimization of the 2D Gaussian convolution into two sequential 1D convolutions - a mathematically equivalent but significantly faster operation. At convergence, the model successfully recovered the reduced grain-boundary thermal conductivity that had been obscured in the low-resolution, analytically obtained thermal image. Figures~\ref{fig:epoch_evolve} and~\ref{fig:kappa_evolve} illustrate the evolution of the bulk and grain-boundary thermal conductivities during training.

It should be noted that the interface conductance, $G$, was held constant across the entire structure to match the ground-truth model. This choice ensured close agreement with the FEM data and allowed the study to focus on recovering the thermal conductivity, as the present work is intended as a proof-of-concept.
 
The initial guess for $G$ in the neural network was obtained by averaging the values from analytical inversion, and this estimate remained unchanged throughout the training process. In the FEM-generated datasets, other material properties influencing thermal transport - namely density and heat capacity - were also held constant across the substrate domain. For real heterointerfaces where these properties may vary, potentially producing different conductances between the transducer and the sample, future implementations may benefit from clustering and fitting $G$ on a per-region basis.

\begin{figure}[H]
\centering
\includegraphics[scale=0.5]{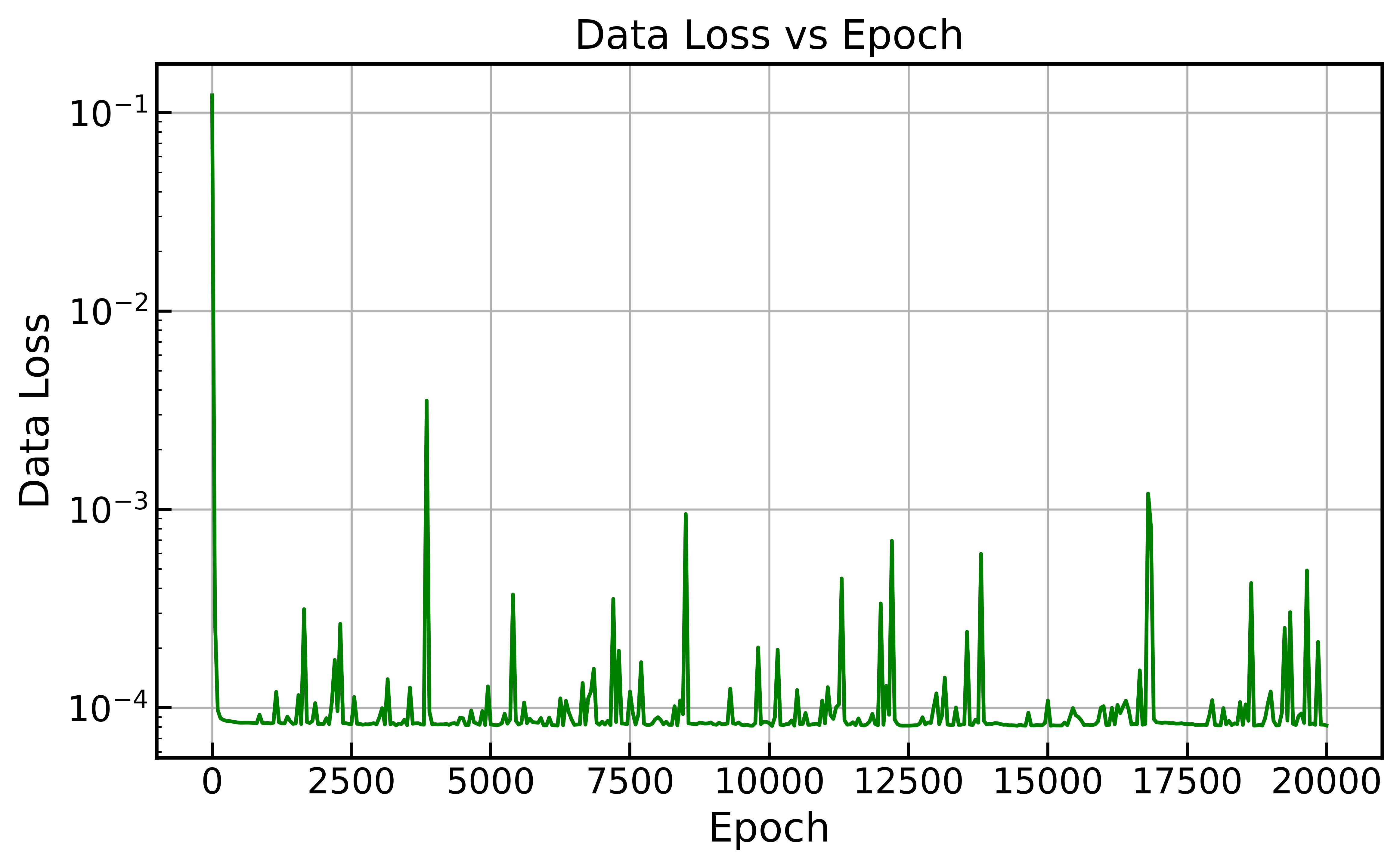}
\caption{Neural Network Data Loss Evolution}
\label{fig:data_evolve}
\end{figure}

\begin{figure}[H]
\centering
\includegraphics[scale=0.5]{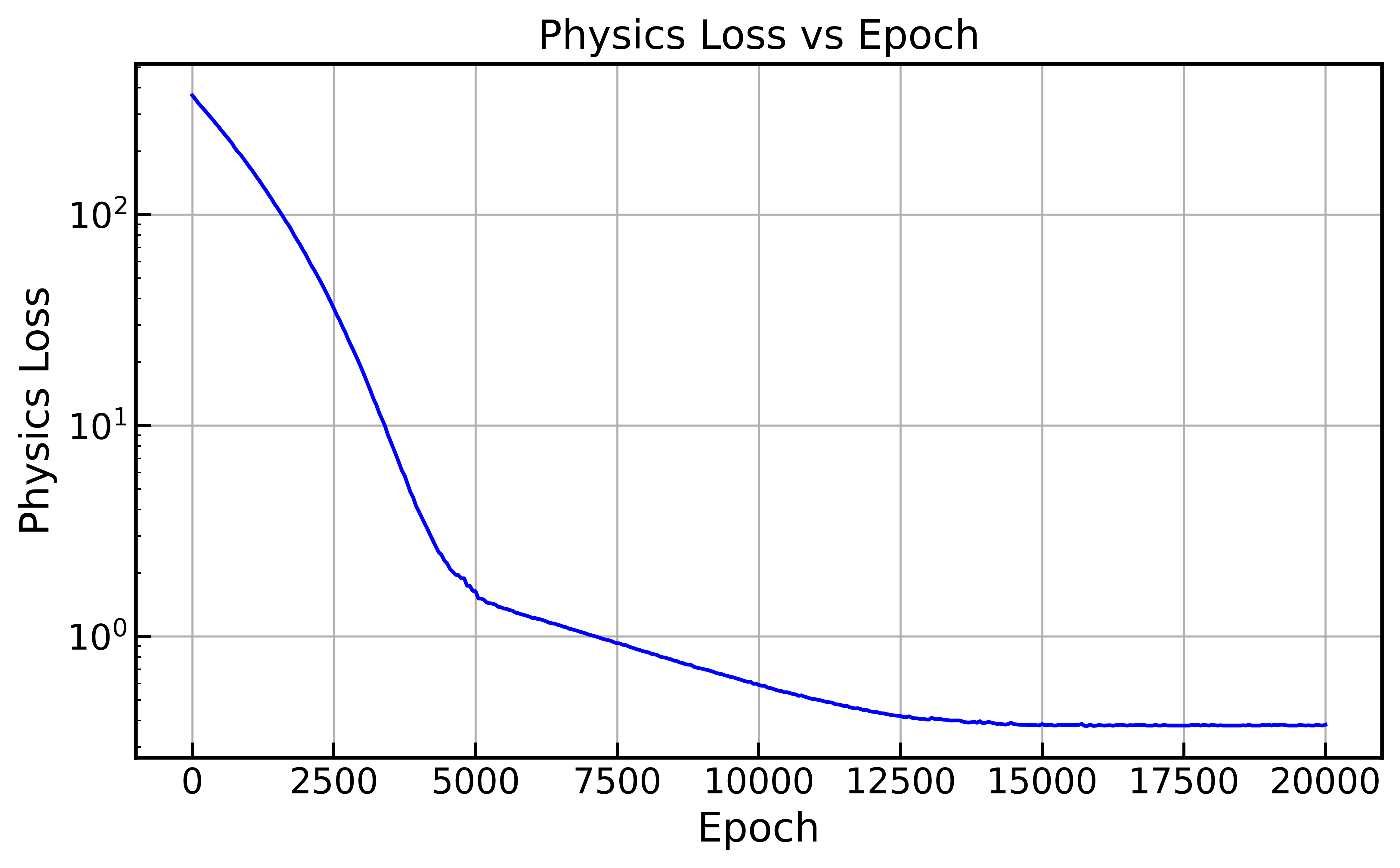}
\caption{Neural Network Physics Loss Evolution}
\label{fig:physics_evolve}
\end{figure}

From Figures \ref{fig:epoch_evolve} and \ref{fig:kappa_evolve}, it is interesting to note how the predicted kappa initially matched the visual trend from the analytically-obtained thermal image (i.e., smooth transition between $130$ and $100$ W/(m.K)) before the physics loss eventually forced a reduction in thermal conductivity at the boundary. Figures \ref{fig:data_evolve} and \ref{fig:physics_evolve} show the evolution of the physics and data losses, respectively.

We observed an almost immediate convergence of the data loss within the early epochs of training, followed by fluctuations as the physics loss began to influence the optimization. In contrast, the physics loss decayed approximately exponentially, with grain boundary features becoming distinguishable after approximately $4,000$ epochs. The process converged at a root-mean-square error (RMSE) of $0.61$~W/(m$\cdot$K), which is close to, but higher than, the $0.22$~W/(m$\cdot$K) RMSE obtained during the determination of $\sigma_1$ and $\sigma_2$. This discrepancy is primarily due to the large error in the grain boundary conductivity (Table~\ref{tab:kappa_comparison}). Sensitivity tests were carried out on different image resolutions other than the $1000 \times 1000$ pixel image presented here, but no consistent trend was observed, with some performing slightly better and others performing considerably worse. This reinforces the need for more robust CNN-learned convolution kernels that account for the structural image resolution in addition to capturing the nuances of FDTR physics which are not captured in the Gaussian convolution.

\begin{table}[H]
\centering
\caption{Comparison of FEM ground truth and learned thermal conductivities.}
\label{tab:kappa_comparison}
\begin{tabular}{lccc}
\hline
\textbf{Parameter} & $\boldsymbol{\kappa_{\text{left}}}$ \textbf{(W/(m$\cdot$K))} & $\boldsymbol{\kappa_{\text{right}}}$ \textbf{(W/(m$\cdot$K))} & $\boldsymbol{\kappa_{\text{GB}}}$ \textbf{(W/(m$\cdot$K))} \\
\hline
FEM ground truth & 130.00 & 100.00 & 75.50 \\
Learned $\kappa$ & 129.56 & 99.59 & 33.74 \\
\% Error         & 0.35\% & 0.41\% & 55.31\% \\
\hline
\end{tabular}
\end{table}

These findings emphasize that while the Gaussian surrogate effectively captures first-order spatial averaging, higher-fidelity convolution kernels - potentially learned through supervised CNNs - are required to ensure resolution invariance and quantitative grain-boundary accuracy.

Furthermore, although the physics loss is more heavily weighted than the data loss (because the thermal conductivity is orders of magnitude larger than the phase), implementations of PINNs usually introduce a loss weighting factor, $\lambda$ (not to be confused with the neuron weights within the neural network, which are handled automatically), to the total loss depending on their specific application:

\begin{equation}
    \mathcal{L}_{PINN} = \lambda_{data}\mathcal{L}_{data} + \lambda_{physics}\mathcal{L}_{physics}
    \label{eq:pinn_loss_mod}
\end{equation}

However, unlike conventional optimization problems where training stops once a specific loss value is reached, in this work the simulation was terminated when each thermal conductivity value converged to a stable value – that is, when further iterations produce negligible change. Consequently, the loss weighting factor has a limited influence on the final outcome, since convergence is governed by the physical stability of the conductivity map rather than a numerical loss threshold. Given the computational efficiency of the model, additional tests with varying weighting factors produced no observable differences in convergence rate or final values, indicating that the formulation is robust to moderate variations in loss weighting.

Overall, these results demonstrate the ability of the model to capture thermal conductivity drops apparent in structural images which are not visually recognizable in lower resolution thermal images.

\newpage
\section{Usage}
\begin{figure}[H]
\centering
\includegraphics[scale=0.5]{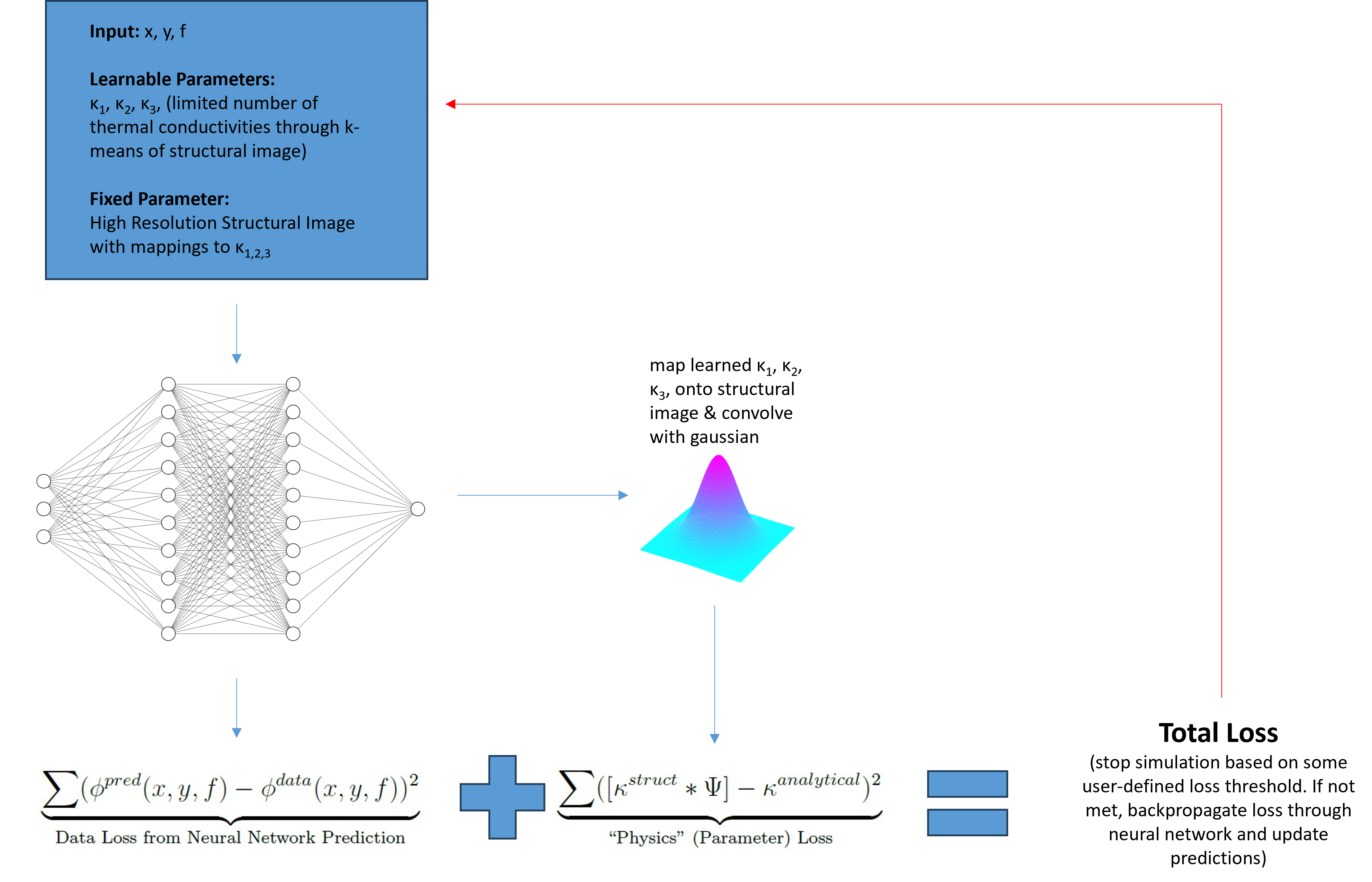}
\caption{\textit{FDTRImageEnhancer} Flow Chart}
\label{fig:flowchart}
\end{figure}

Figure \ref{fig:flowchart} is a flow chart that summarizes how the model works. 

Given phase ($\phi$) data for multiple frequencies saved in \texttt{.txt} files and a structural image (for example, a \texttt{.png} image of structural information), the following steps are recommended.

\begin{itemize}

    \item Pre-process the structural/EBSD image to have a limited number of ``regions" and estimate $\sigma_1$ and $\sigma_2$.
    
    \item Run \texttt{Test\_Data/Structural\_Image\_Clustering.py} to get a file, \texttt{region\_map.npy}

    \item Save/format phase data in \texttt{.txt} files similar to how it is in the \texttt{Test\_Data/} folder.
    
    \item Run an analytical inversion of the phase data to obtain physics-based $\kappa$ and $G$ with \texttt{Test\_Data/Analytical\_Kappa\_Map\_Generation.py}. It will produce two files, \texttt{analytical\_kappa\_map.npy} and \texttt{analytical\_G\_map.npy}. \texttt{kappa\_map.npy} will be compared with the convolution of $\kappa^{pred}$.

    \item Update the file directories at the top of \texttt{FDTRImageEnhancer.py} and run. Observe $\kappa$ evolution and terminate upon convergence of all $\kappa$ values.
    
\end{itemize}

The code is modular and documented to enable modification and extension to other use cases.

\newpage
\section{Conclusion \& Recommendations for Future Work}

In this work, we presented \textit{FDTRImageEnhancer}, an open-source framework that integrates physics-based modeling with machine learning tools to recover grain boundary (GB) thermal conductivity from Frequency Domain ThermoReflectance (FDTR) phase data. This study is intended as a methodological proof-of-concept, demonstrating the feasibility of combining a convolution model with microstructure-aware deep learning. While our initial tests used synthetic data, the results consistently recovered qualitative GB conductivity drops that were visually obscured in analytically inverted maps, whereas bulk values were recovered with $<0.5$\% error. 

\begin{figure}[H]
\centering
\includegraphics[scale=0.5]{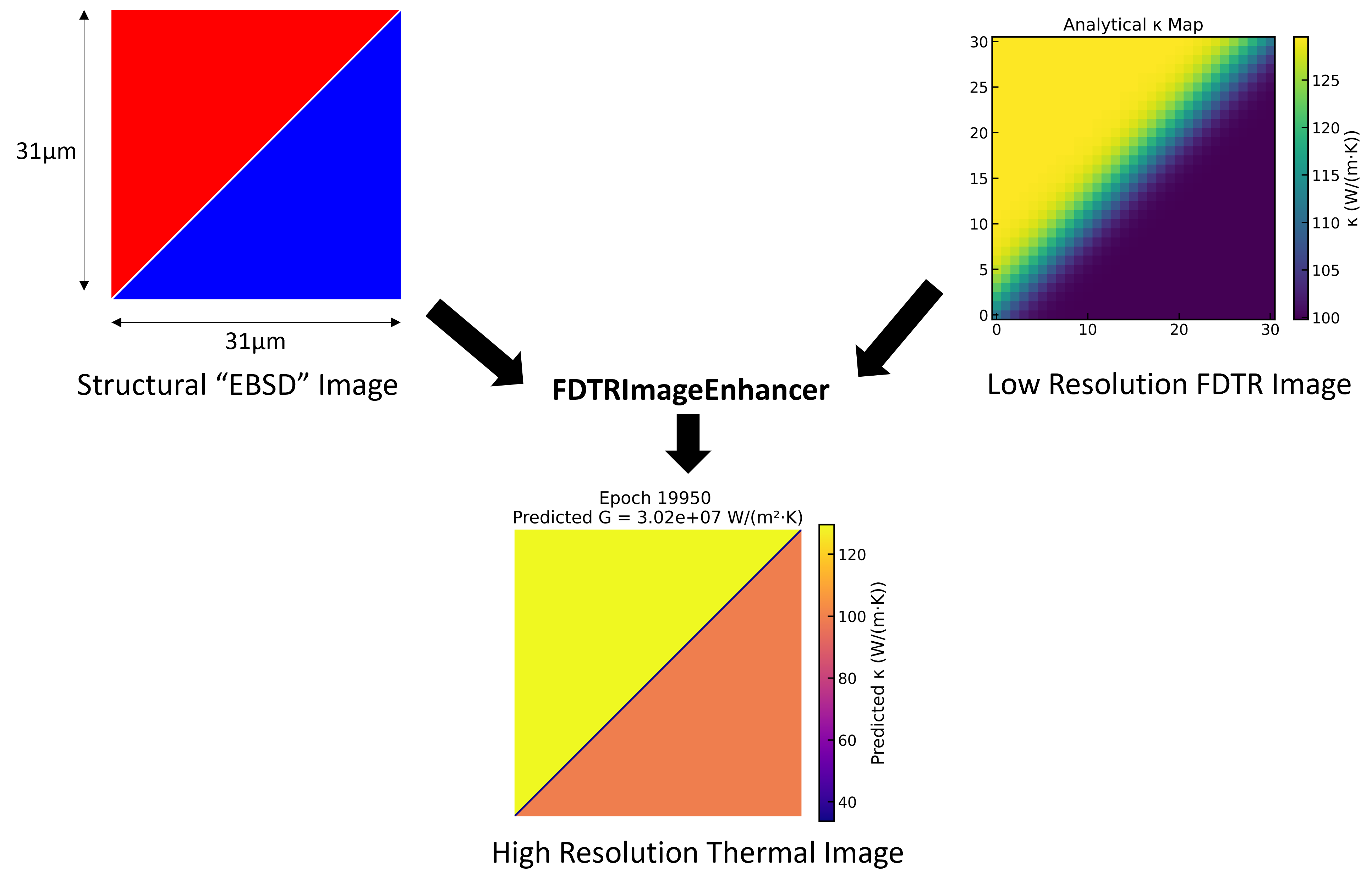}
\caption{\textit{FDTRImageEnhancer} Summary}
\label{fig:fdtr_enhance}
\end{figure}

Looking ahead, we see open directions in which this framework could be developed further by the broader research community:

\begin{itemize}
    \item \textbf{Richer convolution models:} The Gaussian convolution employed in this study is an intentionally simplified representation of FDTR physics - designed to capture the dominant spatial averaging effects while omitting finer-scale interactions between the laser beams and the material. Future work could explore supervised, physics-constrained convolutional neural networks (CNNs) trained on diverse FDTR datasets to better capture these spatial nuances. Such models would be inherently robust to variations in structural image resolution and could enable near-instant ``virtual FDTR experiments'' on high-resolution structural maps. Progress in this direction may offer a computational alternative - or complement - to experimental efforts aimed at improving beam resolution, such as nano-FDTR.

    \item \textbf{Transfer to other domains:} The Gaussian-constrained inverse modeling concept may be valuable in other inverse problems, such as extracting localized strain magnitudes from fracture images in concrete or other quasi-brittle materials.
\end{itemize}

In summary, this work presents a reproducible and extensible starting point for linking high-resolution structural imaging with lower-resolution thermal data in a physically interpretable manner. We hope \textit{FDTRImageEnhancer} will serve as both a demonstration of feasibility and an open invitation for further innovation at the intersection of thermal transport, microstructure analysis, and machine learning.

\newpage
\section{Appendix}

\subsection{Mathematical Justification for Gaussian Convolution}

\subsubsection{Hankel transform review}

The zero-order Hankel transform of a radial function $f(r)$ is defined as:

\begin{equation}
    \bar{F}(k) = \mathcal{H}_0(f) = \int_o^\infty f(r) J_0(kr)r ~dr
    \label{eq:hankel}
\end{equation}

And the inverse:

\begin{equation}
    f(r) = \mathcal{H}_0^{-1} (\bar{F}) = \int_0^\infty \bar{F}(k) J_0(kr)k ~dk
    \label{eq:hankel_inv}
\end{equation}

This transform is the radial analog of the Fourier transform and is used in spatial systems with cylindrical symmetry. 

In real space, the radial convolution of $f(r)$ with some function $y(r)$ is:

\begin{equation}
    [f * y] (r) = \int_0^\infty f(r')y(|r - r'|)r'~dr
    \label{eq:real_conv}
\end{equation}

That is, the convolution at some point $r$ is the integral (sum) of the product of $f$ at that point and $y$ ``shifted" to be centered at the same point. Similar to the Fourier transform, a convolution in real space is akin to a product in Hankel space. That is,

\begin{equation}
    \mathcal{H}_0(f * y) = \mathcal{H}_0(f) \cdot \mathcal{H}_0(y)
    \label{eq:conv_product}
\end{equation}

Inverting both sides, we have:

\begin{equation}
    f * y = \mathcal{H}_0^{-1} (\mathcal{H}_0(f) \cdot \mathcal{H}_0(y))
    \label{eq:invert}
\end{equation}

That is,

\begin{equation}
    f * y = \int_0^\infty \bar{F}(k)  \bar{Y}(k) J_0(kr)k ~dk
    \label{eq:final_conv_def}
\end{equation}

\subsubsection{Application to FDTR}
Let us now assume some function of space, $f(r)$, and define our Gaussian convolution function as $y(r)$. We can evaluate the convolution of $f$ and $y$ at the beam center, that is, $r = 0$, since each scan location is treated as its own origin in the beam-centered coordinate system. This allows $J_0(kr) = 1$, and the convolution becomes: 

\begin{equation}
    [f * y] (0) = \int_0^\infty \bar{F}(k)  \bar{Y}(k) k ~dk
    \label{eq:conv_at_0}
\end{equation}

Since $J_0(0) = 1$. Now recall the complex-valued temperature response from Equation \ref{eq:h_thermal_response}:

\begin{equation}
    \hat{H}(\omega) = \int_0^\infty  \bar{\hat{\mathsf{G}}}(k, \omega)  \bar{\hat{P}}(k, \omega) \bar{\hat{S}}(k, \omega)k ~dk
    \label{eq:h_thermal_response_appendix}
\end{equation}

Relating these expressions to our Gaussian convolution model, we can say that:

\begin{itemize}
    \item $\bar{Y}(k) \approx \bar{\hat{P}}(k, \omega) \bar{\hat{S}}(k, \omega)$. This is because the Hankel transform of a Gaussian is still a Gaussian (and the product of two Gaussians is also a Gaussian!), just with different parameters. Thus, the smoothing parameters $\sigma_1$ and $\sigma_2$ in our Gaussian convolution expression can be roughly correlated to pump and probe laser spot size.

    \item $\bar{F}(k) \approx \bar{\hat{\mathsf{G}}}(k, \omega)$. This implies that $f(r)$ is the Green’s function solution in real space.
\end{itemize}

We further note that while the Green's function solution is dependent on several properties including the heat capacity, material density, etc., it appears to be dominated by the thermal conductivity, as suggested by the close similarity between the phase maps and the recovered thermal conductivity distributions.

Overall, although not strictly one-to-one, this derivation provides a reasonable mathematical justification for the Gaussian convolution approximation, and is intended to serve as a starting point, or ``initial condition'' for a convolutional neural network (CNN) kernel that fully captures FDTR physics.

\newpage
\subsection{FDTR Finite Element Implementation Details}

This section lays out the full implementation details of the Finite Element Model used to generate synthetic FDTR data in this work. These simulations were carried out using the open-source MOOSE Framework \cite{giudicelli2024} and run on a High Performance Computing (HPC) cluster. To improve the computational efficiency of the simulations, they were implemented in the frequency domain, since the experimental method similarly depends solely on the harmonic components of the heating and sensing laser profiles. Before carrying out any simulations, the model was rigorously validated with the integral-transform based analytical solution \cite{schmidt2008} which has been extensively used on experimental data \cite{isotta2024}, and the computed phase values matched with high accuracy (up to 4 decimal places).

The simulation domain consists of a $320 \times 160 \times 40.09 $ $\mu$m box, with the extended $x$ and $y$ directions intended to simulate infinite domains in those directions, and the $z$ profile modeling a semi-infinite domain. The $31 \times 31 ~\mu$m region used for the thermal conductivity images was thus centered around the origin, with a virtual FDTR experiment carried out in $1 ~\mu$m increments in both the $x$ and $y$ directions.

\begin{figure}[H]
\centering
\includegraphics[scale=0.45]{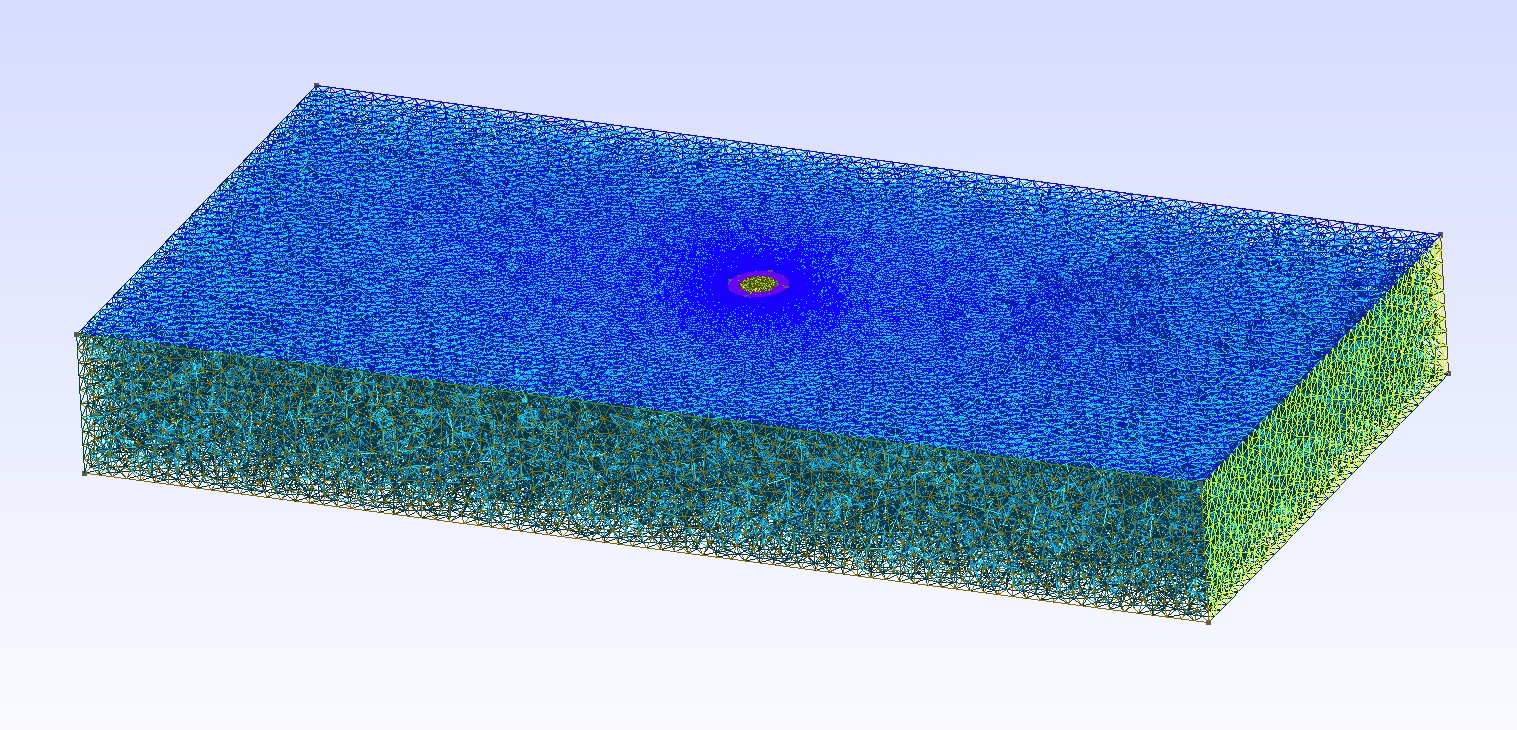}
\caption{FDTR Finite Element Mesh}
\label{fig:fdtr_mesh}
\end{figure}

In the $z$ direction the transducer is assumed to be $90$ nm thick, with material properties for Gold (Au) applied from $0 < z < 0.09$ $\mu$m. Silicon (Si) material properties are used for substrate, i.e the rest of the domain from $-40 < z < 0$ $\mu$m.

The minimum element size is $4$ $\mu$m, which is used for the bulk of the domain. However, closer to the applied heat source, the elements within the transducer have a side length of $0.045$ $\mu$m, and within the substrate the elements are $0.075$ $\mu$m large for $0 < r < 4$ $\mu$m, and $0.2$ $\mu$m large for $4 < r < 8$ $\mu$m, where $r = \sqrt{x^2 + y^2}$. The element sizes smoothly transition in between, ensuring proper resolution of the temperature profile, as the penetration depth is roughly $5$ $\mu$m based on the material properties provided in Section \ref{sec:sim_params}. For each scan location, the mesh refinement placement was updated accordingly. The equations governing the simulations are now presented in the following subsection.

\subsubsection{Frequency Domain Equation Setup}

\paragraph{Fourier Heat Equation}
Since these simulations were run at frequencies within the diffusive regime for heat transport in silicon (i.e $\leq 10$ MHz), the Fourier Heat Equation is assumed to apply, i.e:

\begin{equation}
        \kappa \nabla^2 T - \rho c_p \frac{\partial T}{\partial t} = 0
    \label{eq:append_1}
\end{equation}

where $\kappa$ is the thermal conductivity, $\rho$ is the density, $c_p$ is the heat capacity, and $T$ is the temperature profile. Taking the temporal Fourier Transform, i.e from $T(\mathbf{x}, t)$ to $\hat{T}(\mathbf{x}, \omega)$, this equation becomes:

\begin{equation}
        \kappa \nabla^2 \hat{T} - i \omega \rho c_p \hat{T} = 0
    \label{eq:append_2}
\end{equation}

where $i$ is the imaginary number such that $i = \sqrt{-1}$.

\paragraph{Boundary Condition - Applied Flux}
In the time domain, the applied flux can be modeled as:

\begin{equation}
        q(\mathbf{r}, t) = -\mathbf{n} \cdot (\kappa \nabla T) = \text{Pump}(\mathbf{r}) \times (1 - \cos{(2 \pi f t)})
    \label{eq:append_3}
\end{equation}

Where $\text{Pump}(\mathbf{r})$ is the expression for the shape of the heating profile (regular Gaussian, super-Gaussian ring, etc.) multiplied by a sinusoidal function defined in such a way that the temperature never falls below zero, making it physically realistic. The term $f$ is the heating frequency. In the frequency domain, however, we are only concerned with the harmonic portion of this profile. Therefore, the applied flux becomes:

\begin{equation}
        q(\mathbf{r}, \omega) = -\mathbf{n} \cdot (\kappa \nabla \hat{T}) = \text{Pump}(\mathbf{r}) \times \delta(\omega - 2\pi f)
    \label{eq:append_4}
\end{equation}

All other boundaries are assumed to be zero-flux boundaries.

\paragraph{Interface Condition}
The interface condition in the real space uses the standard continuum approximation, that is,

\begin{equation}
        \mathbf{n} \cdot (\kappa \nabla T^+) = -\frac{1}{R} (T^- - T^+)
    \label{eq:append_5}
\end{equation}

and

\begin{equation}
        \mathbf{n} \cdot (\kappa \nabla T^-) = -\frac{1}{R} (T^+ - T^-)
    \label{eq:append_6}
\end{equation}

where the $+$  and $-$ superscripts denote either side of the interface, thereby enforcing continuity in the flux. In the Fourier space, these expression are simply replaced by their frequency domain counterparts, i.e:

\begin{equation}
        \mathbf{n} \cdot (\kappa \nabla \hat{T}^+) = -\frac{1}{R} (\hat{T}^- - \hat{T}^+)
    \label{eq:append_7}
\end{equation}

and

\begin{equation}
        \mathbf{n} \cdot (\kappa \nabla \hat{T}^-) = -\frac{1}{R} (\hat{T}^+ - \hat{T}^-)
    \label{eq:append_8}
\end{equation}

\paragraph{Initial Conditions}
The initial conditions in the real space assume the material starts at room temperature. However, in the Fourier space, these conditions are not relevant.

\paragraph{Post-Processor - Probe Laser}
The sensing, or probe laser reads the temperature increase integrated over the heating surface, and the phase is computed through the difference between the temperature profile and the heating profile. In the frequency domain, by replacing $T$ with $\hat{T}$, it reads the complex temperature response detailed in Equation \ref{eq:h_thermal_response}. The phase is then computed from the imaginary and real parts as described in Equation \ref{eq:phase}.

\subsubsection{Finite Element Discretization}
\label{sec:fem_fdtr}
The Finite Element Implementation of the governing equations goes as follows:

\begin{itemize}
    \item Decompose $\hat{T}(\mathbf{x}, \omega)$ into imaginary and real parts, i.e $\hat{U}(\mathbf{x}, \omega) + i\hat{V}(\mathbf{x}, \omega)$.

    \item Rewrite heat equation as a system of two equations. The interface condition largely remains unchanged, with the temperature simply replaced by the real and imaginary parts.

    \item The boundary condition (applied heat pump) is fully real, hence it is only applied to $\hat{U}(\mathbf{x}, \omega)$.
\end{itemize}

This section will focus on the Finite Element discretization of the heat equation, since the coupling between the imaginary and real parts of the temperature primarily occurs here. Starting from the frequency-domain heat equation:

\begin{equation}
        \kappa \nabla^2 \hat{T} - i \omega \rho c_p \hat{T} = 0
    \label{eq:append_9}
\end{equation}

Substituting $\hat{T} = \hat{U} + i\hat{V}$, we can separate the equation into its real and imaginary parts:

\begin{equation}
        \kappa \nabla^2 \hat{U} + \omega \rho c_p \hat{V} = 0~~~~~\text{(REAL)}
    \label{eq:append_10}
\end{equation}

\begin{equation}
        \kappa \nabla^2 \hat{V} - \omega \rho c_p \hat{U} = 0~~~~~\text{(IMAGINARY)}
    \label{eq:append_11}
\end{equation}

\paragraph{Weak Form: Real Part}
Starting with the real part of the heat equation, we attempt to get the ``weak form" for the finite element implementation. To do so, we multiply by a test function $\delta u$ and integrate over the domain where equation applies, which we shall call $\Omega$. In tensor notation, this becomes:

\begin{equation}
        \int_\Omega [\kappa \hat{U}_{,ii}\delta u + \omega \rho c_p \hat{V} \delta u]~d\Omega = 0
    \label{eq:append_12}
\end{equation}

From the product rule, we know $(\kappa \hat{U}_{,i} \delta u)_{,i} = \kappa \hat{U}_{,ii}\delta u + \kappa\hat{U}_{,i}\delta u_{,i}$. Substituting in Equation \ref{eq:append_12}:

\begin{equation}
        \int_\Omega [(\kappa \hat{U}_{,i} \delta u)_{,i} - \kappa\hat{U}_{,i}\delta u_{,i} + \omega \rho c_p \hat{V} \delta u]~d\Omega = 0
    \label{eq:append_13}
\end{equation}

We then apply the divergence theorem on the $(\kappa \hat{U}_{,i} \delta u)_{,i}$ to apply the boundary conditions on the 2D surfaces, with the rest of the equation considered the ``kernel" i.e the part that applies to the 3D part of the domain. This is the weak form of the governing equation.

\begin{equation}
        \underbrace{\int_{\partial \Omega} \kappa\hat{U}_{,i} n_{i} \delta u ~d\Omega}_{\text{BC}} - \underbrace{\int_\Omega [\kappa\hat{U}_{,i}\delta u_{,i} - \omega \rho c_p \hat{V} \delta u]~d\Omega}_{\text{Kernel}} = 0
    \label{eq:append_14}
\end{equation}

\paragraph{Weak Form: Imaginary Part}
We obtain the weak form of the imaginary part through a similar process.

\begin{equation}
        -\underbrace{\int_{\partial \Omega} \kappa\hat{V}_{,i} n_{i} \delta v ~d\Omega}_{\text{BC}} + \underbrace{\int_\Omega [\kappa\hat{V}_{,i}\delta v_{,i} + \omega \rho c_p \hat{U} \delta v]~d\Omega}_{\text{Kernel}} = 0
    \label{eq:append_15}
\end{equation}

\newpage
\subsection{Simulation Properties}
\label{sec:sim_params}

Silicon material properties were used for this simulation due to its popularity for use in semiconducting applications.

\begin{table}[H]
\centering
\caption{Simulation and Material Parameters}
\begin{tabular}{|l|l|l|}
\hline
\textbf{Parameter} & \textbf{Symbol / Value} & \textbf{Units} \\
\hline
\multicolumn{3}{|c|}{\textbf{Geometry}} \\
\hline
Transducer thickness (Au) & $9 \times 10^{-8}$ & m \\
Pump radius & $1.53 \times 10^{-6}$ & m \\
Probe radius & $1.34 \times 10^{-6}$ & m \\
\hline
\multicolumn{3}{|c|}{\textbf{Experimental Parameters}} \\
\hline
Absorbed pump power ($Q_0 A$) & $0.01$ & W \\
Au-Si boundary conductance & $3 \times 10^7$ & W/(m$^2\cdot$K) \\
Frequencies sampled & 1, 2, 4, 6, 8, 10 & MHz \\
\hline
\multicolumn{3}{|c|}{\textbf{Material Properties}} \\
\hline
\multicolumn{3}{|l|}{\emph{Silicon (Si) - Sample}} \\
\hline
Bulk thermal conductivity & $130$ & W/(m$\cdot$K) \\
Grain boundary thermal conductivity & $56.52$ & W/(m$\cdot$K) \\
Density ($\rho_{\text{Si}}$) & $2329$ & kg/m$^3$ \\
Heat capacity ($c_p^{\text{Si}}$) & $689.1$ & J/(kg$\cdot$K) \\
\hline
\multicolumn{3}{|l|}{\emph{Gold (Au) - Transducer}} \\
\hline
Bulk thermal conductivity & $215$ & W/(m$\cdot$K) \\
Density ($\rho_{\text{Au}}$) & $19300$ & kg/m$^3$ \\
Heat capacity ($c_p^{\text{Au}}$) & $128.7$ & J/(kg$\cdot$K) \\
\hline
\end{tabular}
\label{tab:parameters}
\end{table}

\section{Data Access Statement}
\label{sec:appendix_code}

The \textit{FDTRImageEnhancer} code and supporting data is available publicly at \texttt{https://github.com/richmondodufisan/FDTRImageEnhancer}.

The Finite Element Model used to generate data is available publicly at \texttt{https://https://github.com/richmondodufisan/purple}.

\section{Conflict of Interest}
The author declares no conflict of interest.



\end{document}